\documentclass{article}
\usepackage{ijcai24}

\usepackage[table,xcdraw]{xcolor}
\usepackage{times}
\usepackage{soul}
\usepackage{url}
\usepackage[hidelinks]{hyperref}
\usepackage[utf8]{inputenc}
\usepackage[small]{caption}
\usepackage{graphicx}
\usepackage{amsmath}
\usepackage{amsthm}
\usepackage{booktabs}
\usepackage{algorithm}
\usepackage{algorithmic}
\usepackage[switch]{lineno}
\usepackage{tabularx}
\usepackage{ltablex}
\usepackage{tabulary}

\newcolumntype{s}{>{\hsize=.3\hsize}X}

\title{Global AI Governance in Healthcare: A Cross-Jurisdictional Regulatory Analysis}

\author{
Attrayee Chakraborty$^1$
\and
Mandar Karhade$^2$ 
\affiliations
$^1$ MS Regulatory Affairs, Northeastern University\\
$^2$ Founder CEO, Citingale\\
\emails
chakraborty.at@northeastern.edu,
mandar.karhade@citingale.com
}

\begin{document}

\maketitle

\begin{abstract}
    Artificial Intelligence (AI) is being adopted across the world and promises a new revolution in healthcare. While AI-enabled medical devices in North America dominate 42.3\% of the global market, the use of AI-enabled medical devices in other countries is still a story waiting to be unfolded. We aim to delve deeper into global regulatory approaches towards AI use in healthcare, with a focus on how common themes are emerging globally. We compare these themes to WHO’s regulatory considerations and principles on ethical use of AI for healthcare applications. Our work seeks to take a global perspective on AI policy by analyzing 14 legal jurisdictions including countries representative of various regions in the world (North America, South America, South East Asia, Middle East, Africa, Australia, and the Asia-Pacific). Our eventual goal is to foster a global conversation on the ethical use of AI in healthcare and the regulations that will guide it. We propose solutions to promote international harmonization of AI regulations and examine the requirements for regulating generative AI, using China and Singapore as examples of countries with well-developed policies in this area.
\end{abstract}

\section{Introduction}
\label{sec:intro}
Artificial intelligence (AI) needs no introduction: it is impacting the healthcare space in unprecedented ways \cite{Rahman2024}. AI is not a single, monolithic technology. Instead, it encompasses diverse subfields, such as machine learning and deep learning, which can be used alone or in combination to create intelligent applications \cite{Bajwa2021}. Machine learning (ML), deep learning and natural language processing have been cited as the most used in diagnosis and treatment recommendations, patient engagement and adherence, and administrative activities \cite{Davenport2019}. Between 1995 and May 2024, the FDA has cleared more than 880 artificial intelligence (AI) medical algorithms. 151 AI-enabled medical devices have been added to the list of approved devices by the FDA as of this year \cite{FDA2024}. AI-enabled medical devices approved by the FDA primarily belong to the medical specialty of radiology \cite{FDA2024}. Key players in the field of AI-enabled devices are located in the US, Canada, and Europe; with North America being the major hub for such devices \cite{Fraser2023} with 42.3\% of the global market share \cite{MaximizeMarketResearch2024}. As of May 2024, 97\% of approved AI/ML-enabled devices (856 out of 882 approved AI/ML-enabled devices) followed the 510(k) pathway in the US showing the prevalence of 510(k) cleared AI/ML-enabled devices in this space \cite{FDA2024}.

AI is being adopted across the world and promises a new revolution in healthcare \cite{Rahman2024}. Legislation is picking up speed, as evidenced by the AI Index Annual Report 2023 which notes 37 measures incorporating AI were signed into law by 2022 as compared to just one in 2016. Additionally, an examination of 81 countries' parliamentary records on AI reveals that since 2016, the number of times AI has been mentioned in international legislative proceedings has increased by almost 6.5 times \cite{AIIndex2023}. This indicates that the global regulatory landscape around AI is highly dynamic at this point in time.

\begin{figure*}
    \centering
    \includegraphics[width=\textwidth]{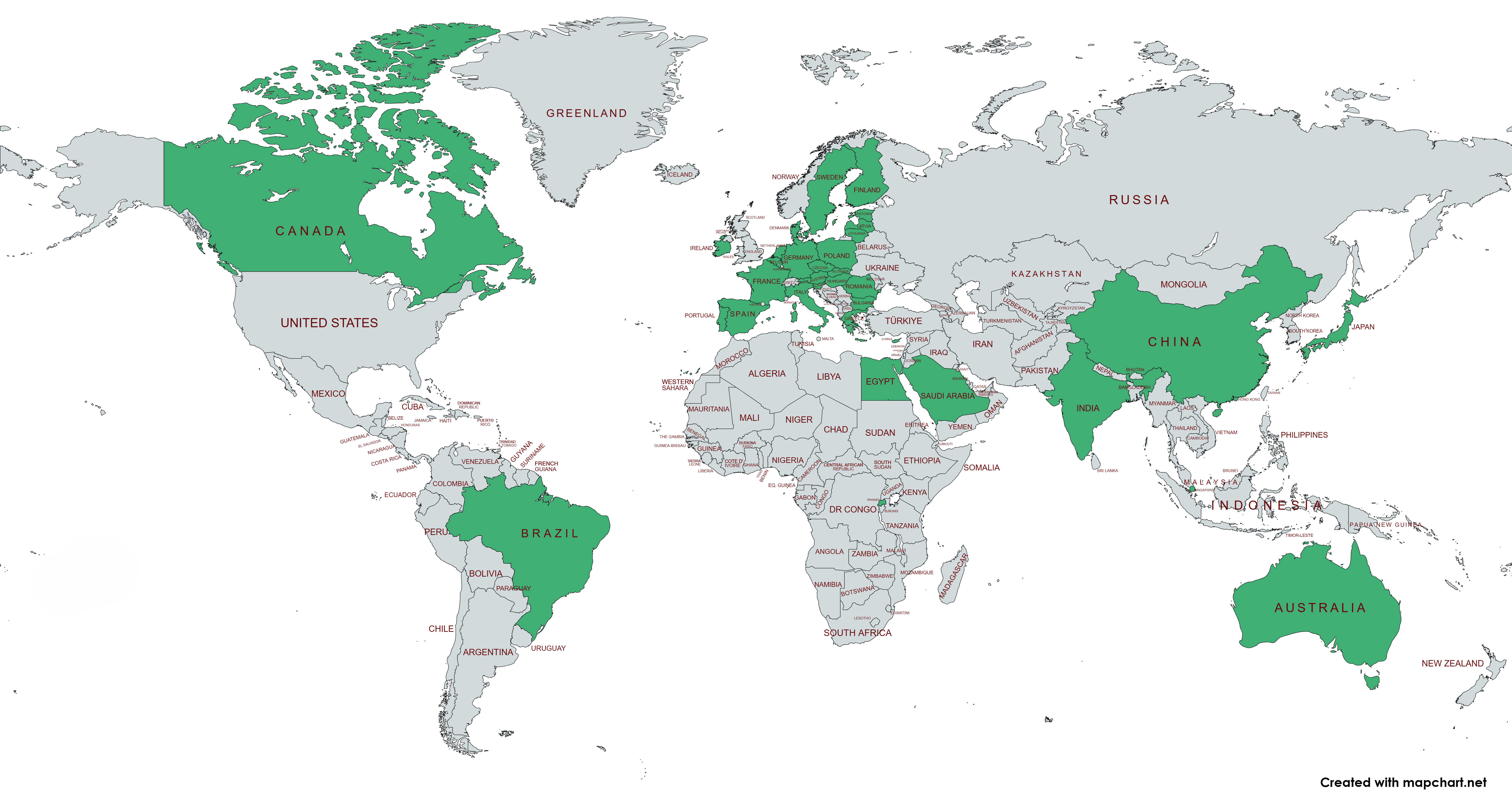}
    \caption{Map depicting areas of legal jurisprudence covered in the scope of this paper (indicated in green)}
    \label{fig:scope_world_map}
\end{figure*}

We have attempted to cover the most recent developments (as of May 2024), however we acknowledge that this paper may not reflect the most recent developments owing to the dynamic nature of AI regulations. Figure \ref{fig:scope_world_map} shows the countries whose AI healthcare regulations are analyzed in this paper.

\section{Related Work}
\label{sec:related_work}
Global AI governance has been well-studied in the legal and technical literature \cite{Schiff2020,daly2020ai,Walter2024}, mainly in the context of principles of AI regulation across countries. Walter introduces the notion of global AI regulation and governance in a sector-agnostic approach, focusing on the socio-economic implications of the rapid advancement of AI technologies and difficulties in establishing effective governance frameworks. Our work, which is in the context of global regulation of AI in healthcare, draws inspiration from Walter to extend sector-agnostic AI governance to sector-specific AI regulation in healthcare. Our work elaborates on AI regulations mentioning healthcare across a global perspective, examining 14 legal jurisdictions from different regions of the world namely, EU, UK, Australia, Canada, Japan, Italy, Brazil, Egypt, Rwanda, Saudi Arabia, Singapore, India, China and Hong Kong. While there is a large amount of literature available on the global regulatory policy and direction with respect to the US government’s approach on AI in healthcare \cite{Wang2019,Chae2020}, there is limited discussion on the status, direction or existing gaps in AI healthcare regulations for other key countries or regions beyond some brief mentions \cite{Tsanni2024}. Murphy et al. \cite{Murphy2021} distinctly highlight the lack of research on AI ethics in Low- or Middle-Income Countries (LMICs) and public health settings. They emphasize the urgent need for further investigation into the ethical implications of AI in these contexts to ensure its ethical development and implementation on a global scale. The scope of our work has been selected keeping in mind this observation, with the larger goal of increasing awareness and representation in conversations relating to the regulation of AI in healthcare. 

There has been a myriad of works on the application of AI in healthcare \cite{Romagnoli2024,Goldberg2024,Jiang2017,Ghosh2024}. Existing literature discusses the requirement of ethical principles in AI governance and provides a high-level discussion of these principles in the context of AI in healthcare \cite{Karimian2022,Giovanola2022,Lehmann2021}. This paper builds on existing literature to analyze laws, regulations, policies, and guidance documents, within our scope, that demonstrate alignment with the WHO’s key principles of ethical AI regulation. 

While applications of generative AI (GenAI) and its governance have been discussed in existing literature \cite{Mesko2023,reddy2024generative,Jindal2024}, our work extends this discussion to country-specific GenAI policies (China and Singapore). This paper also touches upon current legislation on GenAI, given the explosion of large language models (LLMs) like ChatGPT and the growing promise of GenAI to transform clinical workflows, research and medical affairs \cite{Viswa2024}.

\section{Material and Methods}
\label{sec:material_and_methods}
In this work, we have conducted a legal analysis of 25 publicly available laws, guidance documents, and regulations issued in 14 legal jurisdictions (EU, UK, Australia, Canada, Japan, Italy, Brazil, Egypt, Rwanda, Saudi Arabia, Singapore, India, China, and Hong Kong). The choice of nations under the scope of this paper has been made to capture a truly global picture of regulation as elaborated in Section \ref{sec:related_work}\footnote{See comment referencing \cite{Murphy2021} and \cite{Tsanni2024} in Section \ref{sec:related_work} of this paper.}.

We have aimed to incorporate a comprehensive range of regulations related to AI in healthcare. However, currently, the global regulatory landscape predominantly addresses the use of AI in healthcare under the regulatory frameworks established for medical devices, specifically Software as a Medical Device (SaMD) \cite{palaniappan2024global}. Also, most current AI regulations prioritize healthcare but do not provide healthcare-specific regulations \cite{reddy2023navigating}. Therefore, we have analyzed both sector-agnostic generic AI regulations and healthcare-specific AI regulations, mostly in the medical device space. Our definition of sector-agnostic generic AI regulation refers to regulations that govern the use of AI across various sectors and industries, without focusing on the specific applications or risks associated with AI in healthcare specifically. These regulations provide a broad framework for AI governance, addressing general principles and requirements that apply to AI systems regardless of their specific use case. The review also includes national policies in draft or implementation stages, developed by governments, agencies, and standard bodies.

This review considers a mix of four comparative parameters: sector-agnostic generic AI regulations, healthcare-specific AI regulations, non-binding instruments, and binding legal instruments. The regulatory frameworks and guidelines for AI in healthcare across these 14 jurisdictions were identified and downloaded from their respective government healthcare websites for analysis in this review. The search focused on key terms such as regulatory frameworks, legislations, laws, acts, strategies, policies, and guidelines.

We examine quotes and provide references from each of these legal instruments to demonstrate how they align with the WHO's recommended principles for the ethical use of AI in healthcare. The principles include documentation and transparency, risk management, intended use, clinical and analytical validation, data quality, and privacy. This alignment allows us to identify how nations across the world are incorporating WHO guidelines through their strategy, policy, and laws. We identify the legal clauses of jurisprudence tied to WHO’s key principles of ethical AI regulation and reflect its application upon regulations. We also cross-reference publicly available work from international collaboratives and technical focus groups on healthcare-AI regulations to show how their work has and will influence national policies on AI in healthcare.

\section{Global Regulatory Landscape of AI}

\subsection{Definitions}
There is a lack of agreement on what is defined by AI \cite{Krafft2020}. While most nations define specific aspects of AI, such as AI systems \cite{Dwivedi2021}, there is an absence of a clear, widely recognized definition of AI. A notable example is Japan’s acknowledgment of AI as being an ‘abstract’ concept and it is ‘difficult to strictly define the scope of artificial intelligence in a broad sense’ \cite{Ministry2024}, which it rightfully is given that different kinds of AI have become specialized to particular use-cases, an example being GenAI \cite{Kanbach2023}.

\begin{figure*}
    \centering
    \includegraphics[width=\textwidth]{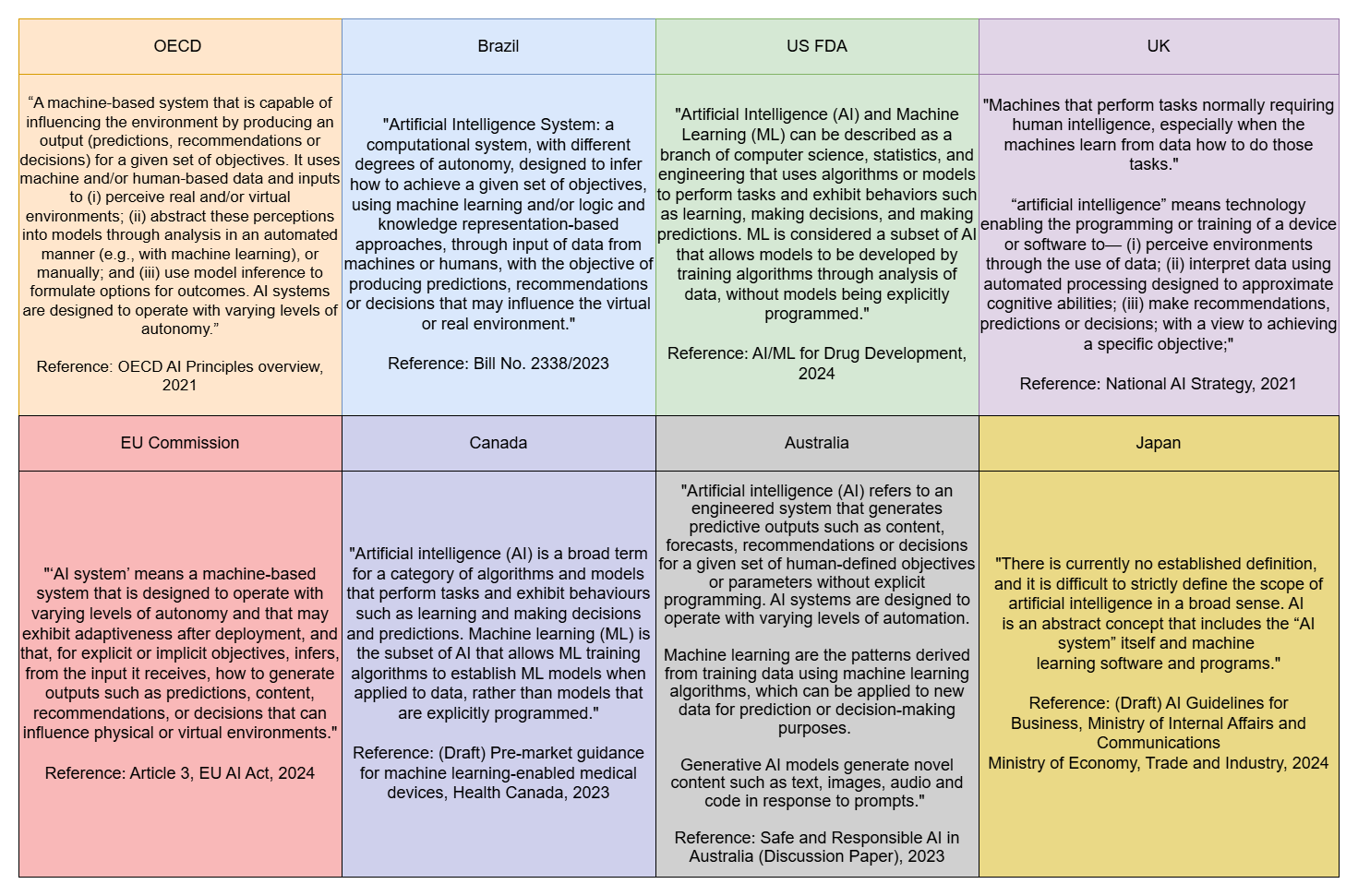}
    \caption{Table representing definitions of AI across nations}
    \label{fig:table_definitions}
\end{figure*}

This ambiguity in AI definition has likely contributed to the field's rapid growth and advancement \cite{Stone2016}. 
Figure \ref{fig:table_definitions} represents how AI is defined in different nations.

\subsection{Common themes in global regulations}
\label{sec:common_themes}
\begin{figure*}
    \centering
    \includegraphics[width=0.9\textwidth]{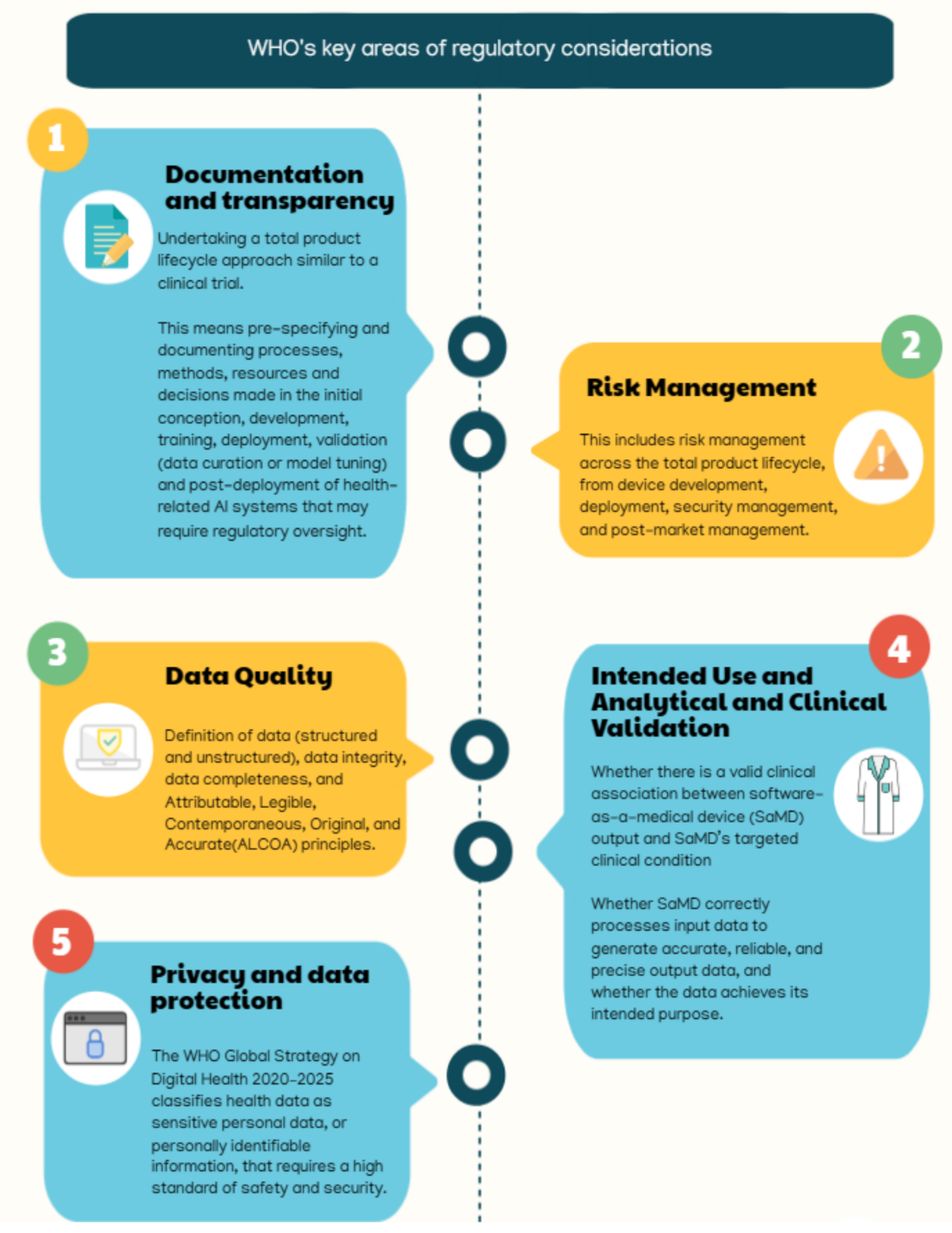}
    \caption{Key regulatory considerations as outlined by the WHO for ethical use of AI in healthcare}
    \label{fig:WHO_principles}
\end{figure*}

The common themes in global AI regulations have been outlined by the OECD \cite{OECD2019}. Existing literature \cite{reddy2023navigating} discusses how general regulations on AI, while providing a broad framework, may not adequately address the specific challenges of AI applications in healthcare. In response to the growing country's need to responsibly manage the rapid rise of AI health technologies, the WHO has responded to the need for frameworks on AI applications in healthcare \cite{WHO2023web} as described in Figure \ref{fig:WHO_principles}.

The principles can be applied to the use of AI in healthcare settings. To illustrate the relevance of these principles, let's consider the clinical setting.

AI models trained on unrepresentative data can perpetuate and worsen existing health disparities due to societal discrimination or small sample sizes \cite{reddy2020governance}. In clinical settings, AI systems must prioritize patient privacy, protect against harm, and ensure patients have control over their data usage \cite{Vayena2018}. Despite the promise of deep learning models in medical imaging and risk prediction, their lack of interpretability and explainability poses significant challenges in healthcare, where transparency is crucial for clinical decision-making \cite{Char2018}. When selecting from multiple algorithms, it is crucial to evaluate risks related to data quality and the suitability of the foundational data to new contexts, such as variations in population and disease patterns \cite{Magrabi2019}. Therefore, evaluation guidelines for AI systems should include assessing and collecting evidence on data quality to prevent unintended consequences and harmful outcomes \cite{Magrabi2019}.

The following sections will elaborate on each of these principles and analyze how different countries are positioning themselves with respect to them.

\subsubsection{Documentation and Transparency}
Transparency ensures that relevant stakeholders receive appropriate information about AI systems \cite{DiazRodriguez2023}. This can be achieved through different levels of transparency, including simulatability (human understanding of the model), decomposability (explaining model behavior and components), and algorithmic transparency (understanding the model's process and output) \cite{BarredoArrieta2020,DiazRodriguez2023}. The ability of AI to learn independently from data poses a challenge when it comes to explaining the decision-making rationale of some AI models \cite{Konigstorfer2022}, posing problems for their application in clinical settings \cite{Smith2021}. Therefore, it is necessary to establish instruments and procedures for confirming that AI applications function as intended and adhere to all applicable laws and regulations \cite{Konigstorfer2022}. Appendix A (Table \ref{tab:1}) explains in detail how laws within the scope of this paper address the WHO's principle of documentation and transparency.

Per our analysis, the EU AI Act\footnote{Chapter III, Article 11, EU AI Act, 2024} is one of the strongest acts declaring the requirement of technical documentation for high-risk AI systems to enable auditing, monitoring and ensuring reproducibility of AI outputs and processes.

A number of other regulations in other countries speak to the same principle (Table \ref{tab:1}). Most laws in AI governance, in healthcare and beyond, mention transparency and explainability as its requirements. However, the definition of transparency varies from ‘communication of appropriate information about an AI system to relevant people’ in the UK \cite{uk_white_paper_2023}  to ‘transparency of governance measures and systems used' in Brazil \cite{brazil_bill_2338_2023}. Transparency is defined in a more structured manner in the context of the healthcare sector by Canada, defining transparency as “the degree to which appropriate and clear information about a device (that could impact risks and patient outcomes) is communicated to stakeholders” \cite{HealthCanada2023}.

\subsubsection{Risk Management}

The National Institute of Standards and Technology (NIST) uses the definition of risk management as mentioned in ISO 31000:2018 for AI systems: “Risk management refers to coordinated activities to direct and control an organization with regard to risk” \cite{NIST2023}. The International Telecommunication Union (ITU) Focus Group on Artificial Intelligence for Health (FG-AI4H) \footnote{The Focus Group on Artificial Intelligence for Health (FG-AI4H) is a partnership of ITU and the World Health Organization (WHO) to establish a standardized assessment framework for the evaluation of AI-based methods for health, diagnosis, triage or treatment decisions.}
elaborates on this thought by its recommendation of “a risk management approach that addresses risks associated with AI systems, such as cybersecurity threats and vulnerabilities, underfitting, algorithmic bias etc.” in the total product lifecycle of an AI system \cite{salathe2018focus}. Appendix A (Table \ref{tab:2}) explains in detail how laws within the scope of this paper address the WHO's principle of risk management.

As per our analysis, risk management is being defined across a spectrum by nations, with prescriptive guidance provided by Brazil on risk classification and risk impact assessment \cite{brazil_bill_2338_2023} and Japan recommending “conducting audits in the AI utilization cycle” \cite{japan_ai_strategy_2022}. Risks linked to cybersecurity and privacy are highlighted by the UK \cite{uk_national_ai_strategy_2021}, while pre- and post-market surveillance is highlighted in Canada’s approach towards medical devices \cite{HealthCanada2023}. Rwanda \cite{rwanda_national_ai_policy_2020} and Egypt \cite{oecd_ai_review_egypt_2023} acknowledge AI risk assessment as a tool for responsible AI, while Singapore \cite{hsa_software_md_2022} and India \cite{icmr_ai_guidelines_2023} have published technical guidance on process controls and change management. Saudi Arabia \cite{sfda_ai_guidelines_2023} emphasizes involvement of a cross-functional team for performing risk management.

\subsubsection{Data Quality}
Data quality is the extent to which a dataset satisfies the needs of the user and is suitable for its intended purpose \cite{Johnson2015}. While data quality issues can impact all modeling efforts, they are particularly problematic in healthcare \cite{Hasan2006}. Data quality issues are particularly challenging in healthcare due to the lack of standardized approaches for describing and handling such issues, the absence of a universal record storage model, the multitude of vocabularies and terminologies used, the inherent complexity of healthcare data, and the ongoing evolution of medical knowledge \cite{Simon2024}.

The ITU FG-AI4H recommends that “developers should consider whether available data are of sufficient quality to support the development of the AI system to achieve the intended purpose \cite{salathe2018focus}. Furthermore, developers should consider deploying rigorous pre-release evaluations for AI systems to ensure that they will not amplify any…biases and errors. Careful design or prompt troubleshooting can help identify data quality issues early and can prevent or mitigate possible resulting harm. Stakeholders should also consider mitigating data quality issues and the associated risks that arise in health-care data, as well as continue to work to create data ecosystems to facilitate the sharing of good-quality data sources” \cite{salathe2018focus}. Appendix A (Table \ref{tab:3}) explains in detail how laws within the scope of this paper address the WHO's principle of data quality.

As per our analysis, we find that Australia exemplifies “data ecosystems” and “sharing of good-quality data sources” through its mention of the healthcare system and national interoperability standards \cite{mq_national_agenda_2023}. 

Japan and Rwanda also propose similar concepts. Japan highlights an important concept of “converting data in a form suitable for AI” and creation of “data economic zones” which will enable the use of AI for healthcare applications \cite{japan_ai_strategy_2022}. Rwanda proposes an implementation plan for availability and accessibility to quality data through indicators such as size of open AI-ready data \cite{rwanda_national_ai_policy_2020}.

While data quality is essential for building accurate AI models, quality culture as an organization influences data management approaches \cite{FDA2019}. UK has a similar approach as it speaks of using a data quality culture, action plans and root cause analysis to address data quality issues at the source \cite{gov.uk2024}. The Framework \cite{gov.uk2024} also speaks of data maturity models and metadata guidance to bring data quality to life. The European Health Data Space (EHDS-TEHDAS) data quality framework recommends more granular mechanisms of data quality management \cite{eurohealthdataspace2024}. Singapore \cite{hsa_software_md_2022}, Hong Kong \cite{mddgovhk2024tr008} and India \cite{icmr_ai_guidelines_2023} also discuss quality of learning and training datasets for accurate validation.

\subsubsection{Intended Use, Analytical and Clinical Validation}
The WHO points to the International Medical Device Regulators Forum (IMDRF)’s definition of clinical evaluation \cite{WHO2023reg}, which consists of valid clinical association, analytical validation, and clinical validation as quoted below:
\begin{itemize}
    \item “Valid clinical association: Is there a valid clinical association between your SaMD output and your SaMD’s targeted clinical condition?

    \item Analytical validation: Does your SaMD correctly process input data to generate accurate, reliable, and precise output data?

    \item Clinical validation: Does use of your SaMD’s accurate, reliable, and precise output data achieve your intended purpose in your target population in the context of clinical care?”
\end{itemize}

On analysis of the jurisprudence within the scope of this paper, we found that national laws in this domain were lacking. While there were technical guidance documents specific to AI applications in healthcare published in Canada \cite{HealthCanada2023}, Singapore \cite{hsa_software_md_2022}, Hong Kong \cite{mddgovhk2024tr008}, Saudi Arabia \cite{sfda_ai_guidelines_2023}, and India \cite{icmr_ai_guidelines_2023}, most of the technical documents by other agencies currently address AI as a subset of software and specific requirements are yet to be updated. Appendix A (Table \ref{tab:4}) explains in detail how laws within the scope of this paper address the WHO's principle of intended use and clinical validation.

The ITU FG-AI4H recommends the use of randomized clinical trials as the gold standard for evaluation of comparative clinical performance, especially for the highest-risk tools or where the highest standard of evidence is required \cite{ITU2022}. It also associates documentation and transparency with validation, mentioning training dataset composition and external analytical validation in an independent dataset. Currently, there are a number of international standards underway, such as ISO/IEC TC215 \cite{isotc215website2024}, IEEE P2802 \cite{standict2024ieeep2802}, and IEC/TC62 PT 63450 \cite{iec2024pt63450} which regulatory guidelines can later reference.

Singapore mentions the type of clinical evidence recommended to support the clinical evaluation process for software and AI-enabled medical devices, such as acceptance limits of testing parameters \cite{hsa_software_md_2022}. Saudi Arabia also notes the absence of internationally aligned frameworks for clinical evaluation of AI/ML enabled medical devices and has gone to reference the IMDRF recommendations, while specifying examples of metrics of clinical validation in intended use environments, such as positive predictive value (PPV) and likelihood ratio negative (LR-) along with mentioning a value  greater than 0.81 as admissible for clinical validation \cite{sfda_ai_guidelines_2023}. 

Another example of specific guidance in this domain is India’s recommendation of the use of Standard Protocol Items: Recommendations for Interventional Trials–Artificial Intelligence (SPIRIT-AI) and Consolidated Standards of Reporting Trials–Artificial Intelligence (CONSORT-AI) as frameworks for designing and running clinical assessment trials related to interventions with AI as a component \cite{icmr_ai_guidelines_2023}. Appendix A (Table \ref{tab:4}) details how the jurisprudence in this paper relates to this principle.

\subsubsection{Privacy and Data Protection}

There have been a number of laws passed in the spirit of privacy and data protection, with the EU GDPR coming into effect in 2018 \cite{general-data-protection-regulation}. The GDPR's data protection by design \footnote{Articles 25 and 32} \cite{general-data-protection-regulation} are being echoed by other nations as well, such as India's proposed data privacy by design policy \cite{IndiaPDP2023}. Privacy impact assessments, a popular approach for proactive privacy risk assessment and mitigation, are frequently included in privacy frameworks. Particular to health data, the European Health Data Space (EHDS) \cite{eurohealthdataspace2024} seeks to foster ownership of healthcare data by individuals and builds further on the GDPR. 

The WHO Global Strategy on Digital Health (2020–2025) \cite{who-global-strategy-digital-health} classifies health data as sensitive personal data, or personally identifiable information, that requires a high standard of safety and security. India’s ICMR guidelines \cite{icmr_ai_guidelines_2023} call out anonymization of data in line with the WHO strategy. It is interesting to note, however, that the anonymization of data does not guarantee privacy, with a study showing how people can be re-identified from an anonymized data collection by providing their zip code, gender, and birthdate \cite{Rocher2019}. 

Singapore \cite{hsa_software_md_2022} emphasizes cybersecurity requirements for connected medical devices, with focus on design controls, test reports, and traceability. Additionally, cybersecurity and privacy go hand in hand, an example being the UK’s ‘Plan for Digital Regulation’ \cite{for_2022} and Saudi Arabia’s ‘Guidance on AI/ML based Medical Devices’ \cite{sfda_ai_guidelines_2023} focusing on infrastructure security. Appendix A (Table \ref{tab:5}) details how the jurisprudence in this paper relates to this principle.

\subsection{Regulations on AI in Healthcare}
\label{sec:regulations_on_ai}

Our analysis of the laws and regulations within the scope of this paper shows that countries are at different stages of developing AI governance frameworks. Most nations are still in the strategy and policy stages of sector-agnostic generic AI regulation, while healthcare agencies in specific countries have been seen to provide healthcare-specific guidance on AI regulation.

\begin{figure*}
    \centering
    \includegraphics[width=\textwidth]{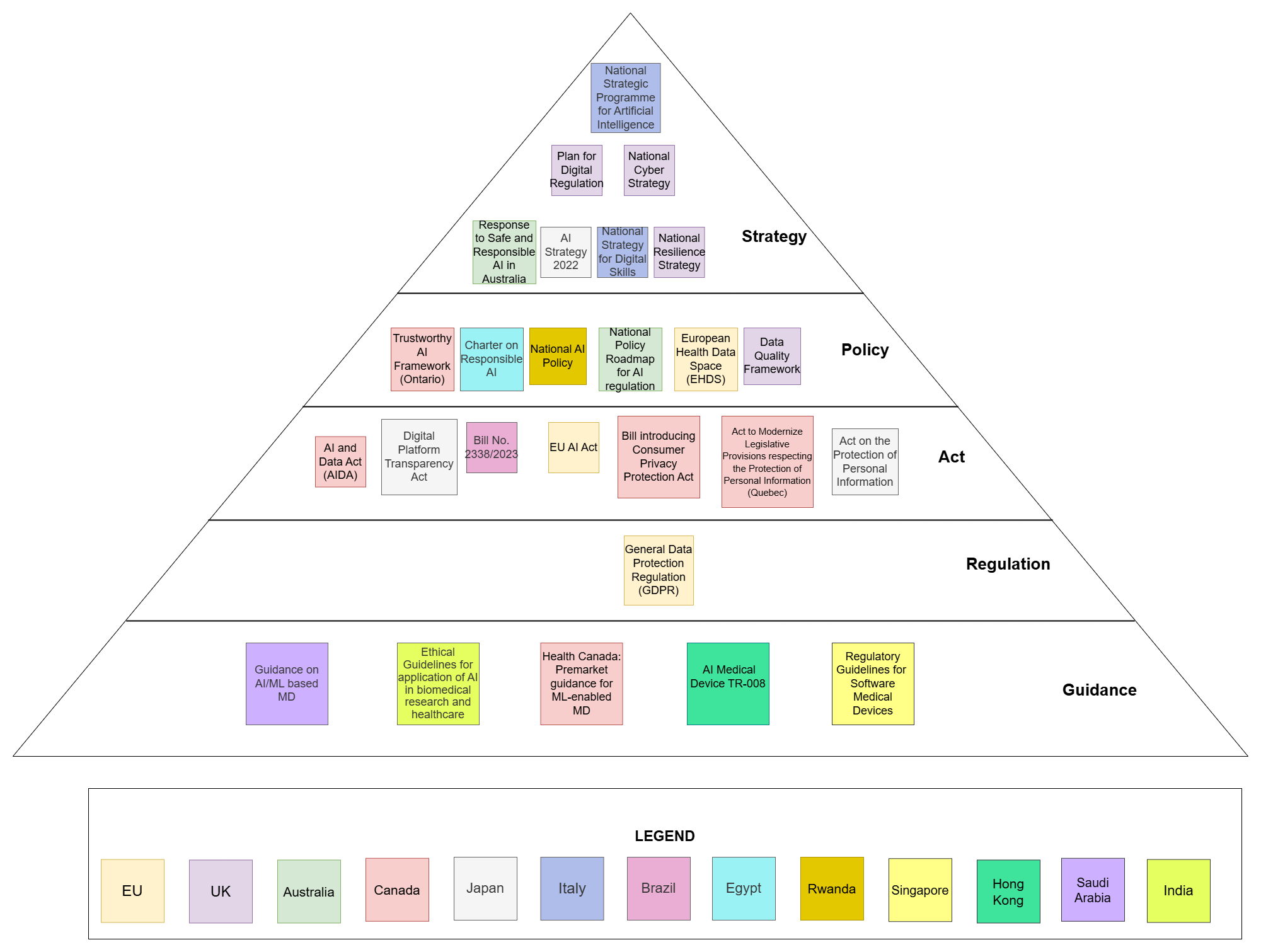}
    \caption{Visual representation of the current AI regulatory landscape in healthcare across 14 jurisdictions}
    \label{fig:law_hierarchy}
\end{figure*}

\begin{figure*}
    \centering
    \includegraphics[width=\textwidth]{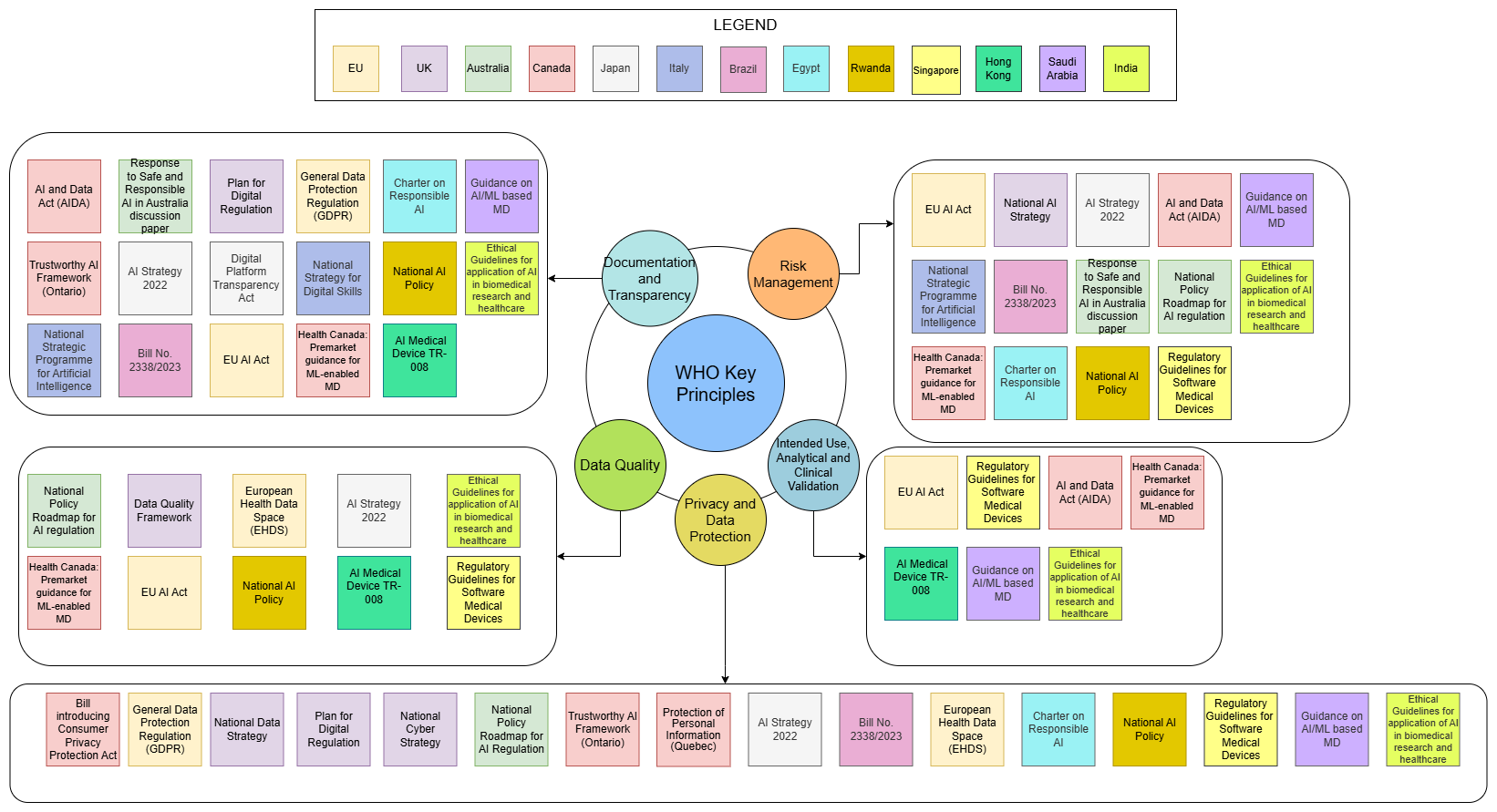}
    \caption{Depiction of laws illustrating WHO’s core principles on ethical AI use in healthcare}
    \label{fig:WHO_diagram}
\end{figure*}

\begin{figure*}
    \centering
    \includegraphics[width=\textwidth]{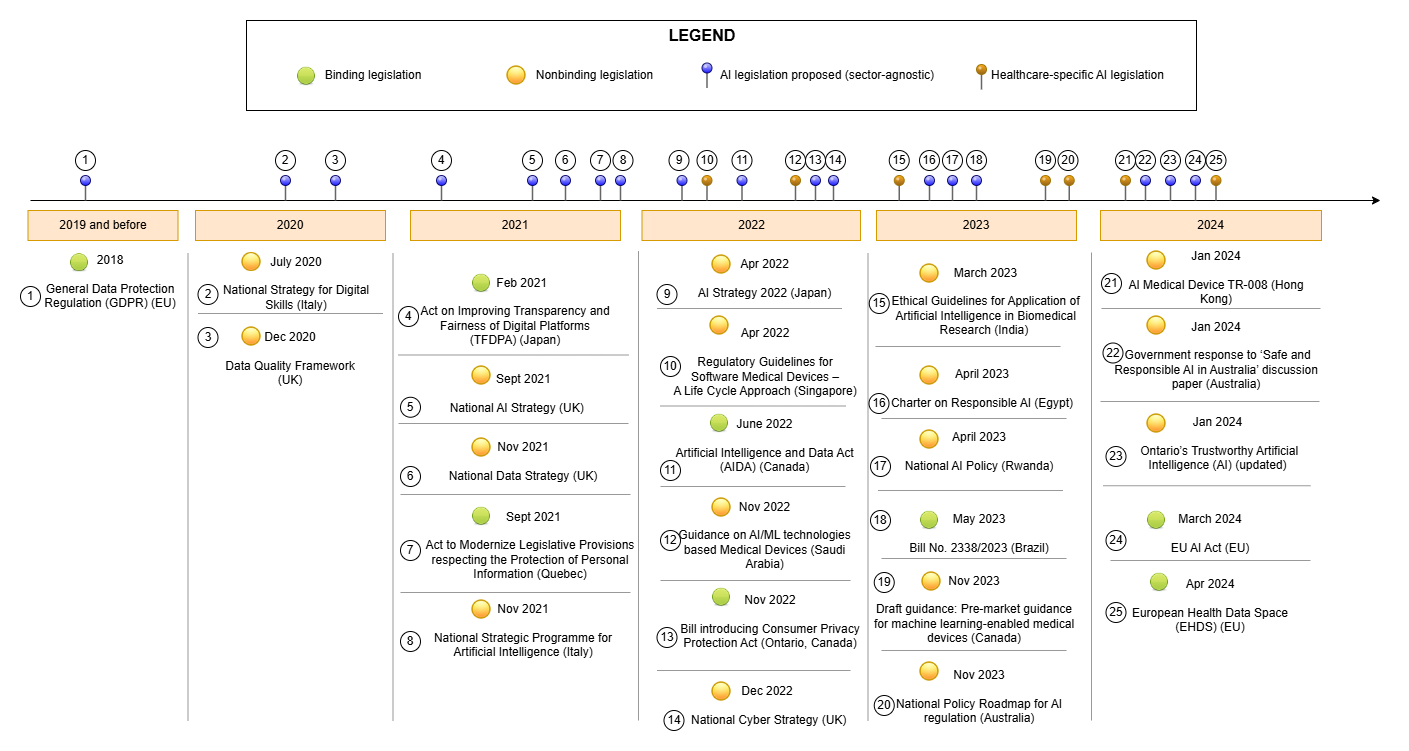}
    \caption{Timeline presenting the evolution of AI legislation across 14 jurisdictions, categorizing each legislation as binding or non-binding and sector-agnostic or healthcare-specific}
    \label{fig:timeline}
\end{figure*}

Figure \ref{fig:law_hierarchy} provides a visual representation of the AI regulatory landscape in healthcare across the jurisdictions analyzed in this paper, illustrating the current state of AI policy development in each region. Figure \ref{fig:timeline} provides a detailed timeline of the laws analyzed in this paper and their relation to AI regulation in healthcare. Figure \ref{fig:WHO_diagram} provides a holistic overview of which jurisprudence in this paper is related to the WHO’s principles for regulation of AI in healthcare \cite{WHO2023reg}(more information presented in Appendix \ref{sec:supp_tables} - Table \ref{tab:1}, Table \ref{tab:2}, Table \ref{tab:3}, Table \ref{tab:4}, and Table \ref{tab:5}).

As per our analysis, most documents discussed in this paper touch upon WHO’s key principles, with internal references to international standards such as ISO and OECD. Most documents discussed have been observed to converge on WHO’s principles.

The most active nations in AI regulation have been the US, EU and China as per recent regulatory reports \cite{Fritz2024}. However, our analysis shows that nations in the Middle East and Southeast Asia are also picking up legislation and policy centered around regulation of AI in healthcare. An example of this is India \cite{ICMR2023}, which calls out ethical principles for AI in healthcare. These principles include acceptable tests for clinical validation of AI used in healthcare and an ethics checklist that covers participant recruitment methods used in training models. Saudi Arabia has also taken a prescriptive approach through elaborating expectations and requirements of AI/ML based device manufacturers, such as clinical evaluation, risk management and quality management systems \cite{SFDA2022}.

African nations are picking up pace on framing policies for AI regulation, with more focus on infrastructure development and user privacy. Rwanda has entered into contracts with digital healthcare providers on AI-powered triage, symptom-checking and cancer detection \cite{AUDA-NEPAD2024}. More details on the prescriptive nature of specific guidance documents, laws, policies and regulations are provided in the supplement (Tables 1, 2, 3, 4 and 5).

We observe that the WHO core principles \cite{WHO2023reg} on AI regulation in healthcare have already been elucidated in various pre-existing standards for medical devices and pharmaceuticals across nations: the principles of transparency, intended use, clinical validation, risk management and privacy have been exhaustively talked about in standards such as ISO 13485:2016 and ISO 14971:2019 published previously. The ITU FG-AI4H approach demarcates AI requirements for medical devices into general, pre-market, and post-market requirements. This approach, which follows the structure of existing total product life cycle approaches for health applications, demonstrates that AI/ML-enabled products have both general requirements (like any other product) and AI-specific requirements that must be considered independently \cite{ITU2022}.

While new AI-specific legal instruments are emerging, many countries are also incorporating AI regulation into existing documents by addressing the additional requirements necessary for AI. For example, Singapore, appended an additional section (Section 9) to its guidance on software as a medical device (SaMD) \cite{hsa_software_md_2022}. This may be an effective stop-gap solution to regulate AI-enabled products in healthcare while more powerful AI laws are being developed.

The convergence of opinion by most nations on regulation of AI used in healthcare is a positive development, given the differing opinions on generic AI regulation. For example, the EU takes a more proactive approach to AI regulation \cite{Stahl2022}, whereas countries like Japan, South Korea and Singapore are mostly prioritizing AI capability development and research \cite{Radu2021}. In contrast, China has adopted a more “top-down” national strategy \cite{Zeng2022}. Italy, on the other hand, has been doubling down on privacy concerns as evidenced through its temporary ban on ChatGPT \cite{Bolici2024}. We believe that the divergent approach regarding generic AI regulation can create a regulatory burden on companies using AI. Comprehensive, healthcare-specific AI regulations are still needed \cite{reddy2020governance}. However, the current reliance on soft-law approaches \cite{palaniappan2024global} allows for the flexibility and adaptability necessary for healthcare regulations to align with WHO guidelines \cite{WHO2023reg}.

While many of the existing AI governance laws are overarching and cover multiple sectors including healthcare, specific focus on regulating AI in healthcare is still a challenge \cite{Simon2024}. This is more relevant with the rise of LLMs in healthcare, such as Med-PaLM, ChatDoctor and ClinicalBERT \cite{Yang2023} which are at the forefront of medical diagnosis, treatment, patient education and clinical documentation.

GenAI application in healthcare is expected to grow at a CAGR of 35.14\% between 2023 and 2032 \cite{PrecedenceResearch2024}, and over two-thirds of US physicians view GenAI as beneficial in healthcare \cite{WoltersKluwer2024}. Regulation of GenAI used in healthcare requires a precise approach \cite{Mesko2023}.

\section{Generative AI: The New Frontier}
\label{sec:genai}
\subsection{GenAI Regulation: Why Do We Need To Regulate It Differently?}
\label{sec:genai_regulation}

Generative AI, unlike traditional AI, uses unsupervised learning and generative models to create entirely new data that resembles training data \cite{Hacker2023}. This makes it extremely vulnerable to hallucinations, bias, and misuse \cite{Nah2023}. Neural network models, which are the core of GenAI, suffer from a lack of transparency and explainability, making it difficult to audit for biases and privacy violations \cite{Salahuddin2022}. While AI governance is picking up speed, regulations surrounding Gen AI may need to be formed keeping these specific differences in mind \cite{Hacker2023}.

The rapid evolution of GenAI makes risk analysis a challenging topic when evaluating business potential, and thus, regulation difficult \cite{Lancet2023}. In current use cases, most GenAI products are trained on structured and unstructured healthcare data containing personal identifiable information \cite{Petrenko2023}. Moreover, while GenAI has the potential to reduce the clinical administrative burden on healthcare workers, inaccurate information can adversely affect patients \cite{Harrer2023}. This underscores the need for clear regulation of GenAI in healthcare settings \cite{Lancet2023}.

\begin{figure}
    \centering
    \includegraphics[width=\linewidth]{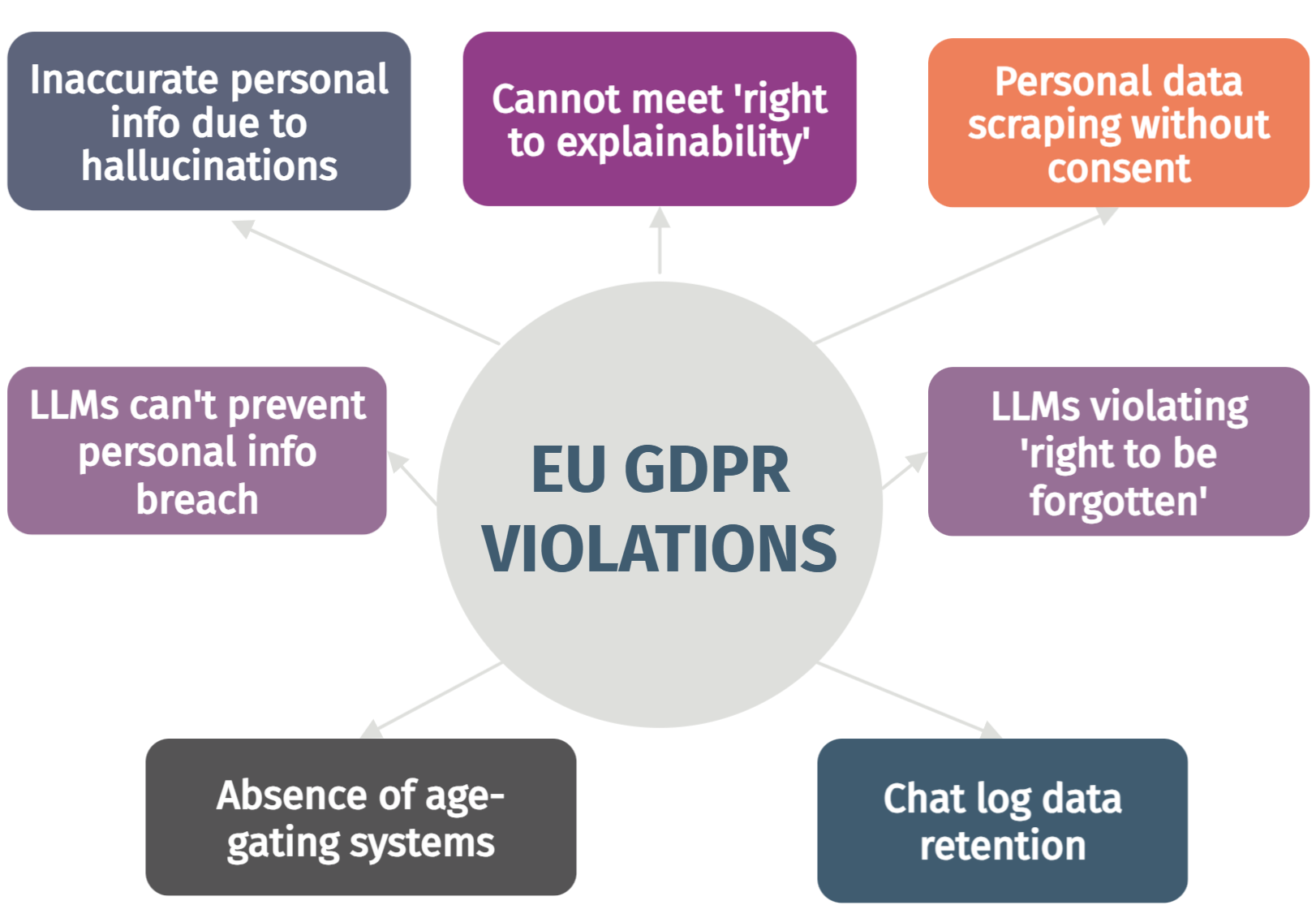}
    \caption{Key instances of how large language models (LLMs) violated EU General Data Protection Regulations (GDPR)}
    \label{fig:llm_gdpr}
\end{figure}

The WHO’s paper on regulatory considerations on artificial intelligence for health \cite{WHO2023reg} highlights how GenAI may already be violating the GDPR as summarized in Figure \ref{fig:llm_gdpr}.

As per the European Data Protection Board (EDPB), several Supervisory Authorities have initiated data protection investigations under Article 58(1)(a) and (b) GDPR against OpenAI (developer of LLM called ChatGPT) in the context of the ChatGPT service \cite{EDPB2024}. There has been a special task force designated for investigating how ChatGPT is positioned with respect to the principles of lawfulness, data collection, fairness, transparency, data accuracy, and subject rights \cite{EDPB2024}. The Australian Alliance for Artificial Intelligence in Healthcare (AAAiH) also recommends communicating “the need for caution in the clinical use of generative AI when it is currently untested or unregulated for clinical settings, including the preparation of clinical documentation.” (Recommendation 4, AAAIH, 2024).

National and global regulatory bodies are struggling to keep pace with the rapid advancements in GenAI, as the technology's trajectory remains uncertain \cite{Lancet2023}. Regulatory mechanisms for GenAI have been proposed advocating three layers of regulation: universal technology-neutral regulation, regulation on high-risk applications of GenAI rather than pre-trained models, and regulation on information access \cite{Hacker2023}. The challenge for regulatory authorities lies in anticipating the full scope of GenAI's evolution and developing comprehensive regulations that address its multifaceted implications \cite{Lancet2023}. We have identified two nations (China and Singapore) with specific GenAI regulations \cite{Luckett2023,Soon2023} as discussed in Section \ref{sec:genai_regulation} as representative examples of GenAI specific regulation.

\subsection{Current Legislation on GenAI}
Currently, we did not come across any legal jurisprudence on GenAI used specifically in healthcare. However, there has been some legal activity on GenAI as a whole. China and Singapore are prominent examples of how GenAI-specific legislation has shaped its legal landscape.

\subsubsection{China}
\begin{figure}
    \centering
    \includegraphics[width=\linewidth]{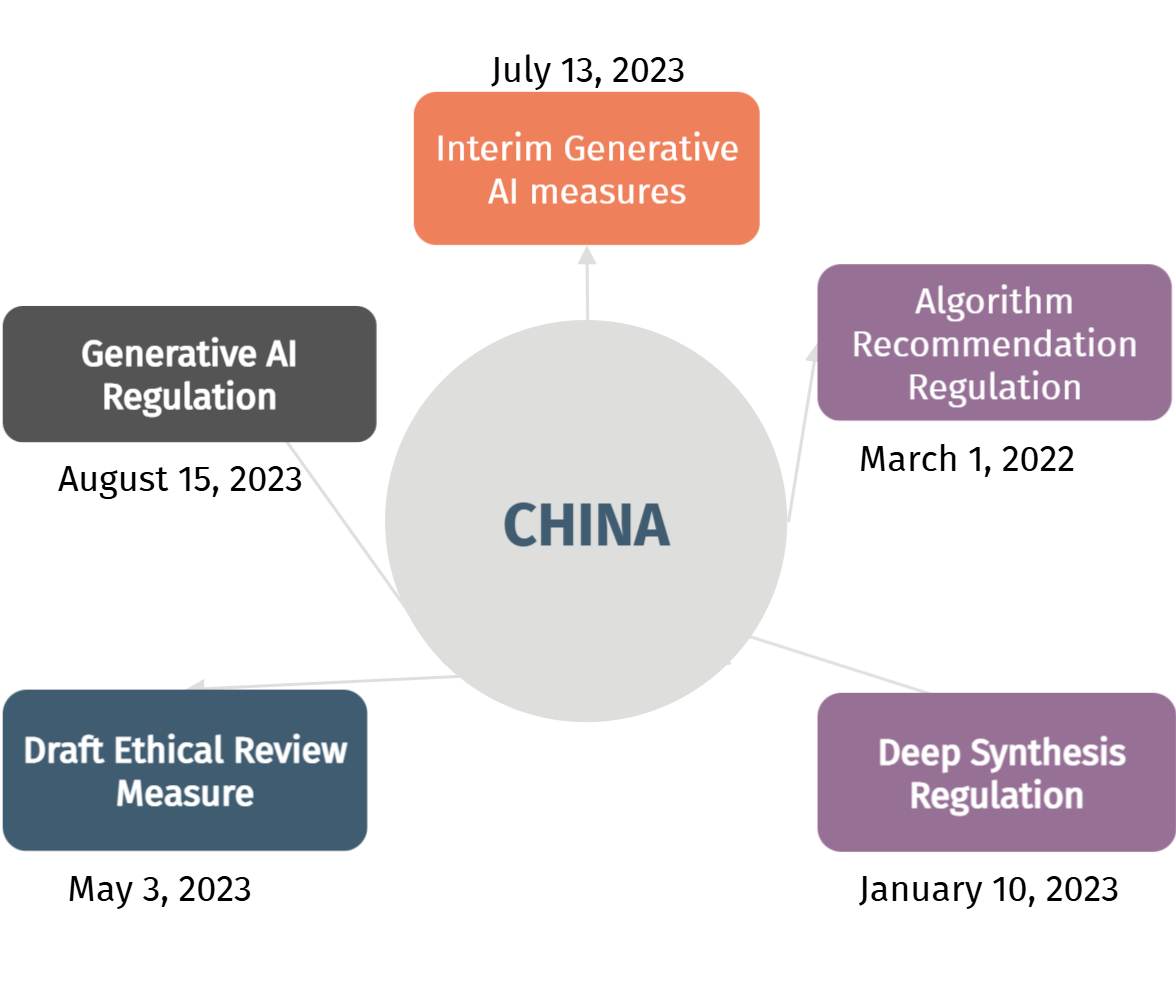}
    \caption{Key regulations on Generative AI (GenAI) in China}
    \label{fig:china}
\end{figure}

The Chinese government has been enacting a number of laws to regulate GenAI. As shown in Figure \ref{fig:china}, the common theme of these laws is their emphasis on regulating data from illegal sources to train models, privacy and security, accountability for content production, tagging GenAI generated content, and complaint management \cite{Wu2023}. A number of other standards have been released early in 2024 such as Draft standards for security specifications on generative AI pre-training and fine-tuning data processing activities (GenAI training data draft standards), Draft standards for security specifications on GenAI data annotation (GenAI annotation draft standards), and Basic security requirements for generative artificial intelligence service (GenAI standards) \cite{Hurcombe2024}. These standards also highlight protection of national security, intellectual property, and protection of individual rights \cite{Hurcombe2024}. It is noteworthy that the Generative AI Measures apply extraterritorially, allowing China to require non-compliant foreign generative AI service providers operating in China to take necessary measures \cite{WANG2024105965}.

\subsubsection{Singapore}
Singapore had released its Model AI Governance Framework in 2019 to lay the groundwork for responsible use of AI. With the rise of GenAI, the AI Verify Foundation and Infocomm Media Development Authority of Singapore (IMDA) of Singapore released its ‘Discussion Paper on Generative AI: Implications for Trust and Governance’ \cite{IMDA2023a}. In response to the discussion paper, AI Verify and IMDA have jointly released the ‘Model AI Governance Framework for Generative AI’ \cite{IMDA2024}. While this Framework focuses on the known topics of data quality, transparency, incident reporting, security, safety and testing, it also focuses on content provenance such as digital watermarking and cryptographic provenance \cite{IMDA2024}. The collaboration has also proposed an ‘initial set of standardized model safety evaluations for LLMs’ including domain-specific tests for medicine \cite{IMDA2023b}.

\section{Results and Conclusion}
In this paper, we analyzed 25 policy, strategy, and guidance based documents, laws and acts centered around AI in healthcare across 14 diverse legal jurisdictions and underscored a global drive towards responsible AI integration within healthcare. While we analyzed most of the specific regulation on AI in healthcare through non-binding instruments (\ref{fig:timeline}), this has both positive and negative consequences. Non-binding approaches offer flexibility and can be easily adapted to the evolving AI landscape. However, their voluntary nature means organizations may choose not to adopt them.

Our findings (Sections \ref{sec:common_themes}, \ref{sec:regulations_on_ai}, and Appendix \ref{sec:supp_tables}) highlight a shared commitment to aligning with the WHO's ethical AI principles, indicating a promising trajectory for the future of AI in healthcare. However, the variability in the specific strategies and the pace of adoption across regions emphasize the need for ongoing international dialogue and cooperation \ref{sec:regulations_on_ai}. 

Most regulations on AI broadly tackle fundamental principles common to most technologies (such as fairness, transparency, bias and privacy). Specific healthcare-centric AI regulations have mostly been found to be proposed by specific healthcare regulatory bodies in the government, such as the FDA (US), Health Canada (Canada), MHRA (UK), ICMR (India) and others (Section \ref{sec:regulations_on_ai}). As explored in Section \ref{sec:common_themes} and Appendixn\ref{sec:supp_tables}, we conclude how existing legislation converges with WHO principles \cite{WHO2023reg}. We have identified how emerging countries are also building requirements as per WHO principles \cite{WHO2023reg} (Section \ref{sec:common_themes}). This approach has met our objective of focusing on global regulation (Section \ref{sec:material_and_methods}) and provides insights beyond existing literature literature (Section \ref{sec:related_work}).

We have also analyzed regulations on 4 comparative parameters: sector-agnostic generic AI regulations, healthcare-specific AI regulations, non-binding instruments and binding legal instruments. We have examined the timeline of evolution of jurisprudence under the scope of this paper for 14 nations \ref{fig:timeline}.

To take a step further, we have discussed two examples of countries framing policies around GenAI as GenAI promises to transform healthcare (Section \ref{sec:genai}).

\section{Future Directions}
We believe that regulations on AI in healthcare can develop as a three-pronged approach:

\subsection*{Collaboration by stakeholders}

We believe that regulatory bodies can refer to deliverables from focus groups such as the International Telecommunication Union Focus Group on Artificial Intelligence for Health (FG-AI4H). This particular group has published considerations for manufacturers and regulators to conduct comprehensive requirements analysis and streamlining conformity assessment procedures for continual product improvement in an iterative and adaptive manner \cite{ITU2022}. Such technical guidance can ensure that specific considerations of AI in healthcare are addressed in regulatory discussions. A number of international standards are under development at the time of writing of this paper, such as ISO/TC 215 (Health informatics) \cite{isotc215website2024}, ISO/IEC AWI TR 18988 (Artificial intelligence — Application of AI (technologies in health informatics) \cite{iso_18988} and ISO /CD TS 24971-2 (Medical devices — Guidance on the application of ISO 14971 Part 2: Machine learning in artificial intelligence) \cite{for_2024}, to name a few which can be referred to by regulators and the healthcare industry.

The evolution of AI regulation with fast-paced changes in technology \cite{DigitalRegulationPlatform2024} can take inspiration from the nature of regulations on drones, which evolved from being an unregulated technology to a highly regulated one in a short timeframe \cite{Fenwick2017}. We can hope that regulators of AI will adapt to the fast-paced nature of AI and develop sector-specific regulations in a short timeframe as well.

By expanding global regulatory alliance and harmonizing requirements, individual nation states can avoid regulatory blind spots, be more prescriptive about the expectations, and increase the speed of well-regulated, safe, and ethical innovation. Including manufacturers in the process of harmonization has also been called out by the WHO as a way to include all stakeholders \cite{WHO2024}. This approach will likely reduce regulatory burden on manufacturers, healthcare systems, and patients by decreasing avoidable variation in the regulatory requirements thereby maximizing their potential benefits for global health while mitigating potential risks.

\subsection*{A possibility of harmonization}

The call for global harmonization of regulations–be it in pharmaceuticals, biologics or medical devices–has been steadily increasing within the industry over the years \cite{LindstromGommers2019}. As highlighted in this paper, the regulation of artificial intelligence (AI) in healthcare remains in its early stages (Section \ref{sec:regulations_on_ai}), presenting a unique opportunity for harmonization. The foundational principles outlined by the WHO \cite{WHO2023reg} offer a promising framework for alignment. An example of this is the alignment of the EU’s new AI Office, along with the UK’s AI Safety Institute, which could potentially interface to lead to a greater degree of global harmonization of AI regulation. Another example is that of a first-of-a-kind international treaty adopted by the Council of Europe (CoE): the Framework Convention on Artificial Intelligence and Human Rights, Democracy and the Rule of Law (Convention) with 46 member states with countries from all over the world being eligible to join it \cite{CouncilofEurope2024}. Harmonization of regulations on AI in healthcare is yet to be seen, especially with the rise of regulatory sandboxes and existing differences in healthcare systems around the world \cite{Leckenby2021,Cancarevic2021}. However, given the diverse approaches of individual countries, ranging from pro-innovation to pro-risk, achieving true harmonization may prove challenging \cite{Thierer2023}.

Experts are already concerned with the divergence in sector-agnostic AI regulation and healthcare-specific regulations, an example being the EU AI Act and EU MDR being described as an “arranged marriage” and “conjoined twins” \cite{RAPS2024}. Even within the AI space, a variety of definitions may lead to greater confusion down the line. Since the regulation of AI is relatively new in the healthcare space, there is still time to harmonize definitions, terminologies, and legislation related to AI in healthcare. The next step in AI regulation would be to issue healthcare-specific regulations and guidance that resolve any inconsistencies between new and existing frameworks.

\subsection*{Risk as the new focus}

As witnessed in the EU AI Act and WHO guidance on ethical use of LLMs in healthcare, a risk-based approach with a focus on accountability in different stages of the value chain of development, deployment, and provision of AI systems is warranted \cite{WHO2024}. The FDA’s recent inclusion of ISO 14971: 2019 as part of its updated Quality Management System Regulations (QMSR) also echoes similar intentions in incorporating risk in systems \cite{Kolton2024}. In our analysis, we note that multiple countries have mentioned risk management and planning as key expectations (Appendix\ref{sec:supp_tables}). 

It is of interest to see how AI validation tools help in converting principles such as risk management to practice, with more than 230 tools for trustworthy AI spanning across the US and UK \cite{Gunashekar2024}. Existing tools specific to healthcare include Aival (for clinical users), Python NLP (for biomedical literature), Google What-If (for analyzing model prediction changes with changes in dataset), and Optical Flow (medical imaging) to name a few \cite{Gunashekar2024}. OECD website can be a great starting point for healthcare regulators to establish acceptable evidence parameters and for industry members to validate AI systems \cite{OECD2024}. The validity of these tools is yet to be seen, with studies revealing that AI auditing tools on the horizon may be questionable in their effectiveness \cite{Graham2020}.


\section*{Ethical Statement}

There are no ethical issues.

\section*{Acknowledgments}

We would like to extend our appreciation to AI-Global Health Initiative (AI-GHI) for providing the platform to connect with diverse stakeholders. Recognized as an FDA Collaborative Community in 2019 and 2021-2024 (active), the AI-GHI leverages their diverse background of stakeholders from medical device, pharmaceutical, biological, hospital, and research sectors to identify the current and future needs of healthcare and provide guidance in how to navigate barriers and risks around the implementation of AI/ML in all of healthcare. Special thanks to Lacey Harbour for her invaluable help with comments on this paper.

AC extends heartfelt gratitude to Susobhan Ghosh for his invaluable assistance in formatting this paper in LaTeX and meticulous review of the formatting.


\bibliographystyle{apalike}
\bibliography{ijcai24}

\newpage
\onecolumn
\appendix
\section{Supplemental Tables}
\label{sec:supp_tables}

\begin{tabularx}{\textwidth}{XXX}
\caption{Documentation and Transparency}\\
\textbf{Law / Regulation / Act / Policy / Guidance} & \textbf{Status} & \textbf{Quote} \\
\hline
\href{https://www.minister.industry.gov.au/ministers/husic/media-releases/action-help-ensure-ai-safe-and-responsible}{Government response to ‘Safe and Responsible AI in Australia’ discussion paper (Australia)} & Consideration on AI Safety Standard & “Transparency – transparency regarding model design and data underpinning AI applications; labelling of AI systems in use and/or watermarking of AI generated content.” \\
\hline
\href{https://www.ontario.ca/page/ontarios-trustworthy-artificial-intelligence-ai-framework}{Ontario’s Trustworthy Artificial Intelligence (AI) Framework (Canada)} & Early stages: requesting feedback & “No AI in secret: This means that we will provide a clear understanding of how and when AI is used.” \\
\hline
\href{https://www.gov.uk/government/publications/digital-regulation-driving-growth-and-unlocking-innovation/plan-for-digital-regulation-developing-an-outcomes-monitoring-framework}{Plan for Digital Regulation (UK)} 
& 
In October 2023, an Outcomes Monitoring Framework was published to track progress against the Plan's objectives using key indicators. 
&
``Keeping the UK safe and secure online: Objectives: Improve users' ability to keep themselves safe online through greater platform transparency and non-legislative support measures.'' 
\\
\hline
\href{https://gdpr-info.eu/recitals/}{General Data Protection Regulation (GDPR) (EU)} & Published and in force & “Recital 58: The Principle of Transparency: The principle of transparency requires that any information addressed to the public or to the data subject be concise, easily accessible and easy to understand, and that clear and plain language and, additionally, where appropriate, visualisation be used.”\\
\hline
\href{https://digital-strategy.ec.europa.eu/en/policies/regulatory-framework-ai}{EU AI Act (EU)} & In March 2024, the European Parliament voted 71-8 to formally adopt the agreed text of the AI Act. Expected to be officially published in May/June 2024. & “The AI Act introduces transparency obligations for all general-purpose AI models to enable a better understanding of these models and additional risk management obligations for very capable and impactful models. These additional obligations include self-assessment and mitigation of systemic risks, reporting of serious incidents, conducting test and model evaluations, as well as cybersecurity requirements.” \\
\hline
\href{https://www8.cao.go.jp/cstp/ai/aistratagy2022en.pdf}{AI Strategy 2022 (Japan)} & AI Strategy 2022 outlined Japan's AI policies as of last year, the government's approach seems to be evolving towards integrating AI initiatives under its broader innovation strategy framework from 2023 onwards, though there are voices advocating for a fresh dedicated national AI strategy as well. & "Part II(3): Overcoming Vulnerabilities Associated with AI and Digitalization -- Establishing Responsible AI and Strengthening Cybersecurity as Cybernetic Resilience: “It is extremely important that the social infrastructure formed by AI and digitization is fair, transparent, operated in a responsible manner, and secure.”
Part III(3): Initiatives to promote implementation in society of AI: (1) Break the black box nature of AI and resolve concerns:“In addition, it is necessary to improve the reliability of AI through initiatives related to Explainable AI (XAI), which breaks the black box nature of AI by enhancing the transparency and accountability of AI processing, and through technological development in the area of integration of cyber security and AI.”\\
\hline
\href{https://www.meti.go.jp/english/policy/mono_info_service/information_economy/digital_platforms/index.html}{Act on Improving Transparency and Fairness of Digital Platforms (TFDPA) (Japan)} & Published and enforced & “The Act requires the specified digital platform providers to disclose terms and conditions and other information, develop procedures and systems to ensure their fairness in a voluntary manner and to submit a report every fiscal year on the overview of measures that they have conducted to which self-assessment results are attached.” \\
\hline
\href{https://repubblicadigitale.gov.it/portale/documents/20122/992735/National+Strategy+for+Digital+Skills.pdf}{National Strategy for Digital Skills (Italy)} & Adopted and implemented & “The Strategy Italy 2025 sets out a clear horizon for "inclusive and sustainable development" as it defines a
course of action that moves towards the challenge of an ethical, inclusive, transparent, and sustainable
innovation for social well-being.”\\
\hline
\href{https://legis.senado.leg.br/sdleg-getter/documento?dm=9347593}{Bill No. 2338/2023 (Brazil)} & In progress & Article 18, Part VII:”It will be up to the competent authority to update the list of
excessive risk or high risk artificial intelligence systems,
identifying new hypotheses, based on at least one of the following
criteria:– low degree of transparency, explainability and auditability
of the artificial intelligence system, which makes its control or supervision difficult.“ (Translated)

Chapter IV, Section I, Article 19:
“Artificial intelligence agents are established
adequate governance structures and internal processes to ensure the security of
systems and compliance with the rights of affected people, under the terms set out
in Chapter II of this Law and the relevant legislation, which will include, at least:
I – transparency measures regarding the use of systems of systems
artificial intelligence in interaction with natural people, which includes the use of
adequate and sufficiently clear human-machine interfaces and
informative;
II – transparency regarding the governance measures governed in the
development and use of the artificial intelligence system by
Organization.”

Article 24: “The impact assessment methodology will contain, at the same time,
least the following steps…transparency measures to the public, especially to
potential users of the system, regarding residual risks, mainly
when it involves a high degree of harm or danger to health or
user safety…”

Article 3: “The development, implementation and use of systems
of artificial intelligence will observe good faith and the following principles:...– transparency, explainability, intelligibility and
auditability.” \\
\hline
\href{https://ised-isde.canada.ca/site/innovation-better-canada/en/artificial-intelligence-and-data-act-aida-companion-document}{AI and Data Act (AIDA) (Canada)} & Introduced in June 2022, proposed amendments in November 2023 & “Transparency means providing the public with appropriate information about how high-impact AI systems are being used.
The information provided should be sufficient to allow the public to understand the capabilities, limitations, and potential impacts of the systems.” \\
\hline
\href{https://www.canada.ca/en/health-canada/services/drugs-health-products/medical-devices/application-information/guidance-documents/pre-market-guidance-machine-learning-enabled-medical-devices.html}{Health Canada: Premarket guidance for ML-enabled MD (Canada)}& Draft & “From our perspective, MLMD lifecycle includes... transparency” \\
\hline
\href{https://read.oecd-ilibrary.org/science-and-technology/oecd-artificial-intelligence-review-of-egypt}{Egyptian Charter on Responsible AI} (Egypt) & Published & “End user right to know when interacting with AI system, ability to challenge AI outcomes, boost awareness and develop pedagogy in AI” \\
\hline
\href{https://www.minict.gov.rw/index.php?eID=dumpFile\&t=f\&f=67550\&token=6195a53203e197efa47592f40ff4aaf24579640e}{National AI Policy (Rwanda)} & Approved & Recommendation No. 8: “By strengthening the capacity 
of regulatory authorities to understand and regulate AI aligned with emerging global 
standards and best practices, we will build transparency and trust with the public.”\\
\hline
\href{https://www.mdd.gov.hk/filemanager/common/mdacs/TR008.pdf}{Medical Device Administrative Control System (MDACS) AI Medical Device TR-008 (Hong Kong)} & Published & Section 4.4: “For AI-MD with CLC, complete information on the learning process including the process controls, verification, on-going model monitoring measures shall be clearly 
presented for review in the application for listing AI-MD.”\\
\hline
\href{https://www.sfda.gov.sa/sites/default/files/2023-01/MDS-G010ML.pdf}{Guidance on AI/ML based Medical Devices (Saudi Arabia)} & Published & Quality Management Systems: “The QMS shall assist the organization to produce a systematic documentation
of the AI/ML and its supporting design and development, including a robust and 
documented configuration and change management process, and identifying its constituent 
parts, to provide a history of changes made to it, and to enable recovery/recreation of past 
versions of the software, i.e., traceability of the AI/ML.” \\
\hline
\href{https://main.icmr.nic.in/sites/default/files/upload_documents/Ethical_Guidelines_AI_Healthcare_2023.pdf}{Ethical Guidelines for application of AI in biomedical research and healthcare (India)} & Published & Section 1.3: Trustworthiness: “Explainable, i.e., the results and interpretations provided by AI-based 
algorithms should be explainable based on scientific plausibility…The end-user must be provided with 
adequate information in a language they can understand to ensure 
that they are not being manipulated by the AI technologies."\\
\hline
\label{tab:1}
\end{tabularx}

\begin{tabularx}{\textwidth}{XXX}
\caption{Risk Management}\\
\textbf{Law / Regulation / Act } & \textbf{Status} & \textbf{Quote} \\
\hline
\href{https://www.minister.industry.gov.au/ministers/husic/media-releases/action-help-ensure-ai-safe-and-responsible}{Government response to ‘Safe and Responsible AI in Australia’ discussion paper (Australia)} & Consideration on AI Safety Standard & “We have heard loud and clear that Australians want stronger guardrails to manage higher-risk AI.” “The Government’s response is targeted towards the use of AI in high-risk settings, where harms could be difficult to reverse, while ensuring that the vast majority of low risk AI use continues to flourish largely unimpeded.”\\
\hline
\href{https://www.mq.edu.au/__data/assets/pdf_file/0005/1281758/AAAiH_NationalAgendaRoadmap_20231122.pdf}{National Policy Roadmap for AI regulation (Australia)} & Published & Recommendations: Point 2: “To ensure AI in healthcare is safe, effective and therefore does not harm patients, it needs to be developed and deployed within a robust risk-based safety framework” Point 1: “To better coordinate and harmonise the responsibilities and activities of those entities responsible for oversight of AI safety, effectiveness, and ethical and security risks, establish a National AI in Healthcare Council.”\\
\hline
\href{https://legis.senado.leg.br/sdleg-getter/documento?dm=9347593}{Bill No. 2338/2023 (Brazil)} & In progress & Chapter III, Risk Categorization: Article 13: “Prior to its placing on the market or use in service, every artificial intelligence system will undergo evaluation preliminary carried out by the supplier to classify its level of risk…There will be registration and documentation of the preliminary assessment carried out by the supplier for liability and accountability purposes in the event that the artificial intelligence system is not classified as risk high…the result of the reclassification identifies the artificial intelligence as high risk, conducting impact assessment algorithmic approach and the adoption of other governance measures provided for in the Chapter IV will be mandatory.” Calls out excessive risk, high risk, Governance Measures for High-Risk Artificial Intelligence Systems (Section II), Algorithmic Impact Assessment (Section III), Codes of Good Practice and Governance (Chapter VI) \\
\hline
\href{https://digital-strategy.ec.europa.eu/en/policies/regulatory-framework-ai}{EU AI Act (EU)} & In March 2024, the European Parliament voted 71-8 to formally adopt the agreed text of the AI Act. Expected to be officially published in May/June 2024. & Article 5 - Prohibited AI practices, Article 6 - Risk Management System for High-Risk AI Systems, Article 7 - Additional Requirements for High-Risk AI Systems, Article 9 - Biometric Categorization Systems, Article 52 - Classification Rules for High-Risk AI Systems, Article 53 - Managing Risks Related to General Purpose AI Systems, Article 61 - AI Systems Presenting Limited Risk, Annex III, Annex VII \\
\hline
\href{https://assets.publishing.service.gov.uk/media/614db4d1e90e077a2cbdf3c4/National_AI_Strategy_-_PDF_version.pdf}{National AI Strategy (UK)} & Current guiding policy framework & “The government is also exploring how privacy enhancing technologies can remove barriers to data sharing by more effectively managing the risks associated with sharing commercially sensitive and personal data.”\\
\hline
\href{https://www8.cao.go.jp/cstp/ai/aistratagy2022en.pdf}{AI Strategy 2022 (Japan)} & AI Strategy 2022 outlined Japan's AI policies as of last year, the government's approach seems to be evolving towards integrating AI initiatives under its broader innovation strategy framework from 2023 onwards, though there are voices advocating for a fresh dedicated national AI strategy as well. & Part II (2) (3): “In order to cope with increasingly complex and sophisticated attacks and the risk of vulnerability that increases as systems become more complex, active consideration should be given to the use of AI, such as information gathering, analysis, support functions, and AI for automation of defense in order to help cyber security analysts make decisions.” Part III (3) (1):“…efforts to realize "Responsible AI" are also expected through initiatives related to the ELSI of AI, such as designing AI with ethical considerations in the first place and conducting audits in the AI utilization cycle.”\\
\hline
\href{https://docs.italia.it/italia/mid/programma-strategico-nazionale-per-intelligenza-artificiale-en-docs/en/bozza/the-strategic-programme-on-artificial-intelligence-anchoring-principles-and-goals.html\#objectives-and-priority-sectors}{National Strategic Programme for Artificial Intelligence (Italy)} & Adopted and Published & Guiding Principles: “On the other hand, the Government is committed to governing AI and mitigating its potential risks, especially to safeguard human rights and ensure an ethical deployment of AI.” \\
\hline
\href{https://ised-isde.canada.ca/site/innovation-better-canada/en/artificial-intelligence-and-data-act-aida-companion-document}{AI and Data Act (AIDA) (Canada)} & Introduced in June 2022, proposed amendments in November 2023 & “The Government has developed a framework intended to ensure the proactive identification and mitigation of risks in order to prevent harms and discriminatory outcomes.” High-impact AI systems: considerations and systems of interest: “The risk-based approach in AIDA, including key definitions and concepts, was designed to reflect and align with evolving international norms in the AI space – including the EU AI Act, the Organization of Economic Co-operation and Development (OECD) AI Principles, and the US National Institute of Standards and Technology (NIST) Risk Management Framework (RMF) – while integrating seamlessly with existing Canadian legal frameworks.” Regulatory Requirements: “AIDA would require that appropriate measures be put in place to identify, assess, and mitigate risks of harm or biased output prior to a high-impact system being made available for use.”\\
\hline
\href{https://www.canada.ca/en/health-canada/services/drugs-health-products/medical-devices/application-information/guidance-documents/pre-market-guidance-machine-learning-enabled-medical-devices.html}{Health Canada: Premarket guidance for ML-enabled MD (Canada)} & Draft & “From our perspective, MLMD lifecycle includes risk management.”\\
\hline
\href{https://read.oecd-ilibrary.org/science-and-technology/oecd-artificial-intelligence-review-of-egypt}{Egyptian Charter on Responsible AI (Egypt)} & Published & “AI risk assessment, reduce harm” \\
\hline
\href{https://www.minict.gov.rw/index.php?eID=dumpFile\&t=f\&f=67550\&token=6195a53203e197efa47592f40ff4aaf24579640e}{National AI Policy (Rwanda)} & Published & “Rwanda’s Guidelines on the Ethical Development and Implementation of Artificial Intelligence, developed by RURA address the range of risks in the AI system lifecycle and considerations for responsible and trustworthy adoption of AI in Rwanda.” \\
\hline
\href{https://www.hsa.gov.sg/docs/default-source/hprg-mdb/guidance-documents-for-medical-devices/regulatory-guidelines-for-software-medical-devices---a-life-cycle-approach_r2-(2022-apr)-pub.pdf}{Regulatory Guidelines for Software Medical Devices – A Life Cycle Approach (Singapore)} & Published & Section 9.2: “If the AI-MD is deployed in a decentralised environment, there should be robust processes in place to address the risks involved in such a decentralised model. Other process controls for consideration includes maintaining traceability, performance monitoring and change management.” \\
\hline
\href{https://www.sfda.gov.sa/sites/default/files/2023-01/MDS-G010ML.pdf}{Guidance on AI/ML based Medical Devices (Saudi Arabia)} & Published & Risk management: “Data scientists should be included in the cross-functional team that perform risk management tasks…There should be a risk management plan that includes...” \\
\hline
\href{https://main.icmr.nic.in/sites/default/files/upload_documents/Ethical_Guidelines_AI_Healthcare_2023.pdf}{Ethical Guidelines for application of AI in biomedical research and healthcare (India)} & Published & Section 1.2: Safety and Risk Minimization: “Some of the risk minimization and safety points are mentioned below…A robust set of control mechanisms is necessary to prevent unintended or deliberate misuse...” \\
\hline
\label{tab:2}
\end{tabularx}

\begin{tabularx}{\textwidth}{XXX}
\caption{Data Quality}\\
\textbf{Law / Regulation / Act } & \textbf{Status} & \textbf{Quote} \\
\hline
\href{https://www.mq.edu.au/__data/assets/pdf_file/0005/1281758/AAAiH_NationalAgendaRoadmap_20231122.pdf}{National Policy Roadmap for AI regulation (Australia)} & Published & Recommendations: Point 10: “Changes may include disclosure to patients that their deidentified patient data is being used to train AI and that clinical recommendations are being based on information provided by AI.” Point 13: “Develop mechanisms to provide industry with ethical and consent-based access to clinical data to support AI development and leverage existing national biomedical data repositories...To maximise national benefit these mechanisms should be based on consistent use of identifiers across the healthcare system and national interoperability standards (e.g. FHIR, SNOMED CT) and be aligned with minimum national datasets and software vendor conformance profiles.” \\
\hline
\href{https://www.gov.uk/government/publications/the-government-data-quality-framework}{Data Quality Framework (UK)} & Guidance, published & Framework provides: "Data quality principles to support organisations to create a data quality culture - A guide to the data lifecycle to help organisations to identify and mitigate potential data quality issues at all stages - Data quality dimensions against which regular assessments of data quality can be made - Data quality action plans, used to identify practical steps to assess data quality and make targeted improvements - Root cause analysis to ensure data quality work addresses issues at source - Metadata guidance to support better use of metadata to communicate and interpret quality - Communicating quality guidance, including suggested approaches for clearly communicating quality to users - An introduction to data maturity models, for those who want to take a holistic approach to assessing and improving data quality.” \\
\hline
\href{https://health.ec.europa.eu/ehealth-digital-health-and-care/european-health-data-space_en}{European Health Data Space (EHDS) (EU}) & Formally approved & Secondary use of health data and the EHDS: “This document identifies several policy options for each barrier, ranging from proposals on improving the clarity of EU data protection law to proposals for improving data quality and interoperability.” Key recommendations of TEHDAS \footnote{The \href{https://tehdas.eu/tehdas1/project/}{TEHDAS1} project (ended in July 2023) developed joint European principles for the secondary use of health data. The work involved 25 countries. The TEHDAS2 joint action started in May 2024 and it will build on the work of previous TEHDAS1.}: “The TEHDAS data quality framework contains the main elements in data quality. These include the steps in the process of preparing data for research and innovation…the European Medicines Agency has used TEHDAS’ data quality work in its efforts to leverage routine data in the real-world evaluation of drugs and medical devices.” \\
\hline
\href{https://www8.cao.go.jp/cstp/ai/aistratagy2022en.pdf}{AI Strategy 2022 (Japan)} & AI Strategy 2022 outlined Japan's AI policies as of last year, the government's approach seems to be evolving towards integrating AI initiatives under its broader innovation strategy framework from 2023 onwards, though there are voices advocating for a fresh dedicated national AI strategy as well. & Part III (3) (2): “In Japan, there is considerable accumulation of high-quality data in each field. Therefore, efforts should be made to enhance data that supports AI utilization by linking and converting these data in a form suitable for AI. With regard to the excellent data base, it is expected that a 'data economic zone' centering on Japan will be constructed by actively engaging in cooperation with other countries.” Part IV (2) (3): Implementation of initiatives that contribute to assurance and confirmation of the quality of collected big data” \\
\hline
\href{https://www.canada.ca/en/health-canada/services/drugs-health-products/medical-devices/application-information/guidance-documents/pre-market-guidance-machine-learning-enabled-medical-devices.html}{Health Canada: Premarket guidance for ML-enabled MD (Canada)} & Draft & “From our perspective, MLMD lifecycle includes…describing the selection and management of data for an MLMD.” \\
\hline
\href{https://digital-strategy.ec.europa.eu/en/policies/regulatory-framework-ai}{EU AI Act (EU)} & In March 2024, the European Parliament voted 71-8 to formally adopt the agreed text of the AI Act. Expected to be officially published in May/June 2024. & Article 7: Data and Data Governance: For high-risk AI systems, this clause mandates using high-quality training, validation and testing data sets that are relevant and representative of the specific geographical, behavioral or functional setting within which the AI system is intended to be used. \\
\hline
\href{https://www.minict.gov.rw/index.php?eID=dumpFile\&t=f\&f=67550\&token=6195a53203e197efa47592f40ff4aaf24579640e}{National AI Policy (Rwanda)} & Published & Implementation Plan Summary: Priority Area 3: Robust data strategy: “Output: Increased availability and access to quality data for training AI models. Indicator: Size (bytes) of open AI-ready data available to the research and innovation community Number of times the open datasets are accessed or downloaded over time.” \\
\hline
\href{https://www.hsa.gov.sg/docs/default-source/hprg-mdb/guidance-documents-for-medical-devices/regulatory-guidelines-for-software-medical-devices---a-life-cycle-approach_r2-(2022-apr)-pub.pdf}{Regulatory Guidelines for Software Medical Devices – A Life Cycle Approach (Singapore)} & Published & Section 9.2: “...For example, there should be appropriate quality checks to ensure that the quality of learning datasets are equivalent to the quality of the original training datasets. There should be validation processes incorporated within the system…” \\
\hline
\href{https://www.mdd.gov.hk/filemanager/common/mdacs/TR008.pdf}{Medical Device Administrative Control System (MDACS) AI Medical Device TR-008 (Hong Kong)} & Published & Section 4.1: Dataset: “The source and size of training, validation and test dataset shall be defined. Information on labelling of datasets, curation, annotation or other steps shall be clearly presented. Description on dataset cleaning and missing data imputation shall also be defined.” \\
\hline
\href{https://main.icmr.nic.in/sites/default/files/upload_documents/Ethical_Guidelines_AI_Healthcare_2023.pdf}{Ethical Guidelines for application of AI in biomedical research and healthcare (India)} & Published & Section 1.6: Optimization of Data Quality: “The manufacturer has the responsibility to eliminate the bias. Demonstration of a bias-free AI technology with the optimum function before a competent authority is mandatory for resuming operations… Training data must not have any sampling bias. Such sampling bias may interfere with data quality and accuracy. Researchers must ensure data quality\\
\hline
\label{tab:3}
\end{tabularx}

\begin{tabularx}{\textwidth}{XXX}
\caption{Intended Use, Analytical and Clinical Validation}\\
\textbf{Law / Regulation / Act } & \textbf{Status} & \textbf{Quote} \\
\hline
\href{https://ised-isde.canada.ca/site/innovation-better-canada/en/artificial-intelligence-and-data-act-aida-companion-document}{AI and Data Act (AIDA) (Canada)} & Proposed amendments, unclear when it will take effect & “Validity means a high-impact AI system performs consistently with intended objectives. Robustness means a high-impact AI system is stable and resilient in a variety of circumstances.” \\
\hline
\href{https://www.canada.ca/en/health-canada/services/drugs-health-products/medical-devices/application-information/guidance-documents/pre-market-guidance-machine-learning-enabled-medical-devices.html}{Health Canada: Premarket guidance for ML-enabled MD (Canada)} & Draft & “From our perspective, MLMD lifecycle includes…design, testing and evaluation, clinical validation, post-market performance monitoring” “The intended use or medical purpose should be made clear in the application…including device function information.” \\
\hline
\href{https://digital-strategy.ec.europa.eu/en/policies/regulatory-framework-ai}{EU AI Act (EU)} & In March 2024, the European Parliament voted 71-8 to formally adopt the agreed text of the AI Act. Expected to be officially published in May/June 2024. & Article 17: Quality management system: “examination, test and validation procedures to be carried out before, during and after the development of the high-risk AI system” Annex IV: Technical documentation referred to in Article 11(1): the validation and testing procedures used” Article 3 (53): ‘real-world testing plan’ means a document that describes the objectives, methodology, geographical, population and temporal scope, monitoring, organisation and conduct of testing in real-world conditions; \\
\hline
\href{https://www.hsa.gov.sg/docs/default-source/hprg-mdb/guidance-documents-for-medical-devices/regulatory-guidelines-for-software-medical-devices---a-life-cycle-approach_r2-(2022-apr)-pub.pdf}{Regulatory Guidelines for Software Medical Devices – A Life Cycle Approach (Singapore)} & Published & Section 3.5 (Clinical evaluation): “The clinical evaluation process establishes that there is a valid clinical association between the software output and the specified clinical condition according to the product owner’s intended use.” “Test protocol and report for verification and validation of the AI-MD, including the acceptance.” \\
\hline
\href{https://www.mdd.gov.hk/filemanager/common/mdacs/TR008.pdf}{Medical Device Administrative Control System (MDACS) AI Medical Device TR-008 (Hong Kong)} & Published & Section 4.1: Performance and Clinical Validation: “Validation and verification test report(s) shall be provided to substantiate such performance claim (e.g. diagnostic sensitivity, diagnostic specificity, accuracy).” \\
\hline
\href{https://www.sfda.gov.sa/sites/default/files/2023-01/MDS-G010ML.pdf}{Guidance on AI/ML based Medical Devices (Saudi Arabia)} & Published & Clinical evaluation: “A manufacturer of AI/ML-based medical devices is expected to provide clinical evidence of the device’s safety, effectiveness and performance before it can be placed on the market.” Analytical validation: “Analytical validation should be done using large independent reference dataset reflecting the intended purpose and the diversity of the intended population and setting.” Intended use: “If the Artificial Intelligence (AI) and Machine Learning (ML) devices are intended by the Product developer to be used for investigation, detection diagnosis, monitoring, treatment, or management of any medical condition, disease, anatomy or physiological process, it will be classified as a medical device subject to SFDA’s regulatory controls.” \\
\hline
\href{https://main.icmr.nic.in/sites/default/files/upload_documents/Ethical_Guidelines_AI_Healthcare_2023.pdf}{Ethical Guidelines for application of AI in biomedical research and healthcare (India)} & Published & Section 1.6: Optimization of data quality: “These inherent problems related to data can be minimized by rigorous clinical validation before any AI-based technology is used in healthcare.” Section 1.10: “AI technology in healthcare must undergo rigorous clinical and field validation before application on patients/participants.” Section 2 of the document, “Guiding Principles for stakeholders involved in development, validation and deployment” describes in detail how AI-based solutions for healthcare must be validated. Section 2.2 describes guiding principles for analytical and clinical validation. \\
\hline
\label{tab:4}
\end{tabularx}

\begin{tabularx}{\textwidth}{XXX}
\caption{Privacy and Data Protection}\\
\textbf{Law / Regulation / Act } & \textbf{Status} & \textbf{Quote} \\
\hline
\href{https://www.gov.uk/government/publications/digital-regulation-driving-growth-and-unlocking-innovation/plan-for-digital-regulation-developing-an-outcomes-monitoring-framework}{Plan for Digital Regulation (UK)}  & In October 2023, an Outcomes Monitoring Framework was published to track progress against the Plan's objectives using key indicators. & “Objectives: Citizens are empowered to be safe online, and trust they are protected from online harms beyond their control; Organisations have the capabilities and resilience to preserve their digital security, and security is factored into new products and services from the outset; The security of UK networks and critical infrastructure is protected.” \\
\hline
\href{https://read.oecd-ilibrary.org/science-and-technology/oecd-artificial-intelligence-review-of-egypt}{Egyptian Charter on Responsible AI} & Published & “Final human determination always takes place especially for sensitive or mission-critical AI applications, data protection and AI risk assessment” \\
\hline
\href{https://www.minict.gov.rw/index.php?eID=dumpFile\&t=f\&f=67550\&token=6195a53203e197efa47592f40ff4aaf24579640e}{National AI Policy (Rwanda)} & Published & Implementation Plan Summary: Priority Area 2: N24: “Publish guidance targeted towards industry and users on how existing privacy legislation fits with cloud computing.” N30: “Enforce Data Protection and Privacy Law.” \\
\hline
\href{https://www.hsa.gov.sg/docs/default-source/hprg-mdb/guidance-documents-for-medical-devices/regulatory-guidelines-for-software-medical-devices---a-life-cycle-approach_r2-(2022-apr)-pub.pdf}{Regulatory Guidelines for Software Medical Devices – A Life Cycle Approach (Singapore)} & Published & Section 8 (cybersecurity) [multiple references] \\
\hline
\href{https://www.sfda.gov.sa/sites/default/files/2023-01/MDS-G010ML.pdf}{Guidance on AI/ML based Medical Devices (Saudi Arabia)} & Published & Risk management: “There should be a risk management plan that includes cybersecurity risks.” \\
\hline
\href{https://main.icmr.nic.in/sites/default/files/upload_documents/Ethical_Guidelines_AI_Healthcare_2023.pdf}{Ethical Guidelines for application of AI in biomedical research and healthcare (India)} & Published & Section 1.4: Data Privacy: “Individual patients’ data should preferably be anonymized unless keeping it in an identifiable format is essential for clinical or research purposes. All algorithms handling data related to patients must ensure appropriate anonymization before any form of data sharing.” \\
\hline
\href{https://health.ec.europa.eu/ehealth-digital-health-and-care/european-health-data-space_en}{European Health Data Space (EHDS)} & Published & “The EHDS will: empower individuals to take control of their health data and facilitate the exchange of data for the delivery of healthcare across the EU.” “In the future, decisions on using health data will be taken by a specialised authority in each country, a health data access body. Access to data would only be allowed for specific purposes. The TEHDAS \footnote{The \href{https://tehdas.eu/tehdas1/project/}{TEHDAS1} project (ended in July 2023) developed joint European principles for the secondary use of health data. The work involved 25 countries. The TEHDAS2 joint action started in May 2024 and it will build on the work of previous TEHDAS1.} project developed a data quality framework which aims to ensure that health data collected across Europe and reused for policymaking, regulation and research is reliable enough and fit for purpose.” \\
\hline
\href{https://ised-isde.canada.ca/site/innovation-better-canada/en/consumer-privacy-protection-act}{Bill introducing Consumer Privacy Protection Act (Canada) }& Consideration stage in the House of Commons committee & “Enhancing Canadians' control and consent…New rules will require transparency on the use of automated systems—such as artificial intelligence—that make decisions and predictions about Canadians….Clearer rules for the handling of de-identified information will facilitate its use for the research and development of innovative goods and services.” \\
\hline
\href{https://gdpr-info.eu/recitals/}{General Data Protection Regulation (GDPR) (EU)} & Published and in force & Recital 1 (Data Protection as a Fundamental Right): “The protection of natural persons in relation to the processing of personal data is a fundamental right” Recital 53 (Processing of Sensitive Data in Health and Social Sector): “Special categories of personal data which merit higher protection should be processed for health-related purposes only where necessary… this Regulation should provide for harmonised conditions for the processing of special categories of personal data concerning health, in respect of specific needs, in particular where the processing of such data is carried out for certain health-related purposes by persons subject to a legal obligation of professional secrecy.” Recital 54 (Processing of Sensitive Data in Public Health Sector): “Such processing of data concerning health for reasons of public interest should not result in personal data being processed for other purposes by third parties…The processing of special categories of personal data may be necessary for reasons of public interest in the areas of public health without consent of the data subject.” \\
\hline
\href{https://www.gov.uk/government/publications/national-data-strategy-mission-1-policy-framework-unlocking-the-value-of-data-across-the-economy/national-data-strategy-mission-1-policy-framework-unlocking-the-value-of-data-across-the-economy\#priorities}{National Data Strategy (UK)} & Published & “A robust regime is already in place: there are categories of data sharing that are not permitted subject to a consent framework and/or can only be done in certain ways to manage those risks, which the government continues to keep under review. Levers to manage this include the Information Commissioner’s Office (ICO)’s data sharing code of practice, the Centre for the Protection of National Infrastructure’s Security-Minded approach to Open and Shared Data, the Official Secrets Act, the Information Management Framework which is currently under development, and other relevant legislation and guidance. The Central Digital and Data Office’s Data Ethics Framework, which is designed to guide public sector use of data, may also inform how organisations in the private and third sector use data.” \\
\hline
\href{https://www.gov.uk/government/publications/national-cyber-strategy-2022/national-cyber-security-strategy-2022\#foreword}{National Cyber Strategy (UK)} & Published & Foreword: “We see this in our response to international health emergencies and in our promotion of Net Zero targets…” Annex B: The NIS Regulations established a new regulatory regime within the UK that requires designated operators of essential services (OESs) and relevant digital service providers (RDSPs) to put in place technical and organisational measures to secure their network and information systems..It applies to sectors…healthcare” Pillar 3, Objective 3: “Our activity in and in relation to cyberspace has enhanced global stability…This will include but not be limited to tackling internet shutdowns, bias in Artificial Intelligence algorithms and increasing online safety.” \\
\hline
\href{https://www.mq.edu.au/__data/assets/pdf_file/0005/1281758/AAAiH_NationalAgendaRoadmap_20231122.pdf}{National Policy Roadmap for AI regulation (Australia)} & Published & Priority Area 1: “Patients must receive safe, effective, and ethical care from AI-enabled healthcare services and be assured sensitive healthcare data are protected from cybersecurity threats, privacy breaches or unauthorised use…Uploading sensitive patient data into a non-medical AI like ChatGPT hosted on United States servers is also problematic from a privacy and consent perspective.” \\
\hline
\href{https://www.ontario.ca/page/ontarios-trustworthy-artificial-intelligence-ai-framework}{Ontario’s Trustworthy Artificial Intelligence (AI) Framework} & Early stages: requesting feedback & Principles for Ethical Use of AI [Beta]: “Data enhanced technologies should be designed and operated in a way throughout their life cycle that respects the rule of law, human rights, civil liberties, and democratic values. These include dignity, autonomy, privacy, data protection, non-discrimination, equality, and fairness.” \\
\hline
\href{https://www.legisquebec.gouv.qc.ca/en/document/cs/p-39.1}{Act to Modernize Legislative Provisions respecting the Protection of Personal Information (Quebec)} & Passed & Division II (8.1): “...any person who collects personal information from the person concerned using technology that includes functions allowing the person concerned to be identified, located or profiled must first inform the person…“Profiling” means the collection and use of personal information to assess certain characteristics of a natural person, in particular for the purpose of analyzing that person’s… health.” Division III, Section 1(18.2): “no information relating to a person’s health may be communicated without the consent of the person concerned unless 100 years have elapsed since the date of the document.” \\
\hline
\href{https://www8.cao.go.jp/cstp/ai/aistratagy2022en.pdf}{AI Strategy 2022 (Japan)} & AI Strategy 2022 outlined Japan's AI policies as of last year, the government's approach seems to be evolving towards integrating AI initiatives under its broader innovation strategy framework from 2023 onwards, though there are voices advocating for a fresh dedicated national AI strategy as well. & Part II (2) (3): “The realization of Responsible AI is a requirement that must be secured in the promotion of digitization. To this end, it will be important to promote further R \& D and implementation in society of a series of technologies such as Explainable AI (XAI) and Federated Learning, which can be learned while protecting privacy and confidential information, as well as to build platforms and to exercise leadership in their operation.“ \\
\hline
\href{https://legis.senado.leg.br/sdleg-getter/documento?dm=9347593}{Bill No. 2338/2023 (Brazil)} & In progress & Article 2, Part VIII: “The development, implementation and use of artificial intelligence systems in Brazil are based on: privacy, data protection and informational self-determination” Article 5, Part VI: “Persons affected by artificial intelligence systems have the following rights, to be exercised in the manner and under the conditions described in this Chapter…privacy, data protection and informational self-determination;” Article 19, Part IV: “Artificial intelligence agents shall establish governance structures and internal processes capable of ensuring the security of the systems and compliance with the rights of affected persons, under the terms provided for in Chapter II of this Law and the relevant legislation, which shall include, at least…legitimacy of data processing in accordance with data protection legislation, including through the adoption of privacy measures by design and by default and the adoption of techniques that minimize the use of personal data.” \\
\hline
\label{tab:5}
\end{tabularx}

\end{document}